\documentclass[twocolumn]{aastex631}

\usepackage{natbib}
\usepackage{graphicx}	
\usepackage{epstopdf}
\usepackage{amsmath}	
\usepackage{amssymb}	
\usepackage{amsfonts}
\usepackage{multirow}

\begin{document}

\title[About the accuracy of {\tt relxill} and {\tt relxill\_nk}]{About the accuracy of the {\tt relxill}/{\tt relxill\_nk} models in view of the next generation of X-ray missions}

\author[0000-0003-2845-1009]{Honghui Liu}
\affiliation{Center for Astronomy and Astrophysics, Center for Field Theory and Particle Physics, and Department of Physics,\\
Fudan University, Shanghai 200438, China}
\affiliation{Institut f\"ur Astronomie und Astrophysik, Eberhard-Karls Universit\"at T\"ubingen, D-72076 T\"ubingen, Germany}

\author{Askar~B.~Abdikamalov}
\affiliation{School of Natural Sciences and Humanities, New Uzbekistan University, Tashkent 100007, Uzbekistan}
\affiliation{Center for Astronomy and Astrophysics, Center for Field Theory and Particle Physics, and Department of Physics,\\
Fudan University, Shanghai 200438, China}
\affiliation{Ulugh Beg Astronomical Institute, Tashkent 100052, Uzbekistan}

\author{Temurbek Mirzaev}
\affiliation{Center for Astronomy and Astrophysics, Center for Field Theory and Particle Physics, and Department of Physics,\\
Fudan University, Shanghai 200438, China}

\author[0000-0002-3180-9502]{Cosimo Bambi}
\affiliation{Center for Astronomy and Astrophysics, Center for Field Theory and Particle Physics, and Department of Physics,\\
Fudan University, Shanghai 200438, China}
\affiliation{School of Natural Sciences and Humanities, New Uzbekistan University, Tashkent 100007, Uzbekistan}

\author{Thomas~Dauser}
\affiliation{Dr. Karl Remeis-Observatory and Erlangen Centre for Astroparticle Physics, D-96049 Bamberg, Germany}

\author{Javier~A.~Garc{\'\i}a}
\affiliation{NASA Goddard Space Flight Center, Greenbelt, MD 20771, USA}
\affiliation{Cahill Center for Astronomy and Astrophysics, California Institute of Technology, Pasadena, CA 91125, United States}

\author{Zuobin Zhang}
\affiliation{Center for Astronomy and Astrophysics, Center for Field Theory and Particle Physics, and Department of Physics,\\
Fudan University, Shanghai 200438, China}

\correspondingauthor{Cosimo Bambi}
\email{bambi@fudan.edu.cn}

\begin{abstract}
X-ray reflection spectroscopy is a powerful tool to study the strong gravity region of black holes. The next generation of astrophysical X-ray missions promises to provide unprecedented high-quality data, which could permit us to get very precise measurements of the properties of the accretion flow and of the spacetime geometry in the strong gravity region around these objects. In this work, we test the accuracy of the relativistic calculations of the reflection model {\tt relxill} and of its extension to non-Kerr spacetimes {\tt relxill\_nk} in view of the next generation of X-ray missions. We simulate simultaneous observations with \textsl{Athena}/X-IFU and LAD of bright Galactic black holes with a precise and accurate ray-tracing code and we fit the simulated data with the latest versions of {\tt relline} and {\tt relline\_nk}. While we always recover the correct input parameters, we find residuals in the fits when the emission from the inner part of the accretion disk is higher. Such residuals disappear if we increase the number of interpolation points on the disk in the integral of the transfer function. We also simulate full reflection spectra and find that the emission angle from the accretion disk should be treated properly in this case.
\end{abstract}



\section{Introduction}\label{sec:Intro}

Accreting black holes show fascinating phenomena that provide valuable insights into the extreme environments around these objects. Among the most intriguing aspects, there are relativistic reflection features that are commonly observed in the X-ray spectra of such systems \citep{1989MNRAS.238..729F, 1995Natur.375..659T, 2007MNRAS.382..194N, 2009ApJ...697..900M}. These features occur when a hot corona illuminates a cold accretion disk \citep{1995MNRAS.277L..11F}, leading to a reflection spectrum that is influenced by the interplay between Compton scattering, absorption, and fluorescent emission, producing prominent fluorescent emission lines, such as the iron K$\alpha$ complex around 6-7~keV, and a Compton hump with a peak typically around 20-30~keV \citep{2005MNRAS.358..211R, 2010ApJ...718..695G}. These reflection spectra appear blurred to a distant observer because of relativistic effects \citep{1989MNRAS.238..729F, 1991ApJ...376...90L, Dauser2010MNRAS.409.1534D, 2017bhlt.book.....B}. The analysis of these blurred reflection features has become a powerful tool for probing the physics and astrophysics in the strong gravity regime of accreting black holes.

In recent years, the development of sophisticated reflection models and new observational facilities have significantly advanced our understanding of these relativistic reflection features \citep{2021SSRv..217...65B}. Currently, X-ray reflection spectroscopy has measured the spin of approximately 40~stellar-mass black holes in X-ray binary systems and around 40~supermassive black holes in active galactic nuclei \citep{2021SSRv..217...65B,2023ApJ...946...19D}. Moreover, it is currently the only mature technique for determining the spins of supermassive black holes, while, for stellar-mass black holes, spin measurements can also be obtained from the thermal spectrum of the disk \citep{1997ApJ...482L.155Z,2014SSRv..183..295M} and from the gravitational wave signal of black hole binaries \citep{2014PhRvL.112y1101V,2016PhRvL.116f1102A}. Additionally, X-ray reflection spectroscopy can be used as an essential tool to test Einstein's theory of General Relativity in the strong field regime, and it currently offers rigorous tests of the Kerr metric around black holes \citep{2018PhRvL.120e1101C,2019ApJ...875...56T,2021ApJ...913...79T,2022ApJ...924...72Z,2022arXiv221005322B}.

In anticipation of high-quality data from future instruments -- such as the X-ray Integral Field Unit (X-IFU) onboard \textsl{Athena} \citep{2013arXiv1306.2307N} and the Large Area Detector (LAD), which is a high throughput instrument that may fly in future and was proposed to be part of the \textsl{eXTP} payload \citep{2016SPIE.9905E..1QZ} -- it is imperative to evaluate the accuracy of existing reflection models. Here we limit our study to test the accuracy of the relativistic calculations, namely of the photon trajectories from the emission points on the accretion disk to the detection point in the flat faraway region and the photon redshifts resulting from the combination of the 4-velocity of the material in the disk and the 4-momentum of the photons at the emission and detection points. We aim to assess the accuracy of the reflection models {\tt relxill} \citep{2013MNRAS.430.1694D,2013ApJ...768..146G,2014ApJ...782...76G}\footnote{The model can be downloaded from \url{http://www.sternwarte.uni-erlangen.de/~dauser/research/relxill/index.html}.} and {\tt relxill\_nk} \citep{Bambi_2017,2019ApJ...878...91A}\footnote{The model can be downloaded from \url{https://github.com/ABHModels/relxill_nk}.} in light of the high-quality data expected from future instruments. We simulate a set of observations with X-IFU and LAD. Our simulation setup includes a range of black hole spin values, inclination angles, and emissivity profile indices to examine the models' performance under various conditions. By fitting the simulated spectra with the {\tt relline} and {\tt relline\_nk} modules from the {\tt relxill} and {\tt relxill\_nk} packages, respectively, we aim to determine the suitability of these models for analyzing the anticipated high-quality data from upcoming X-ray observatories.

The structure of this paper is organized as follows. Section~\ref{sec:ray_tra} provides a comprehensive account of the ray-tracing code used to calculate the iron lines which are the bases for the simulated observations. In Section~\ref{sec:simu}, we describe the methodology used to simulate the X-ray observations with the X-IFU and LAD instruments, including the input models, exposure times, and parameter configurations. We also present the results of our spectral fitting analysis, evaluating the performance of the {\tt relxill} and {\tt relxill\_nk} reflection models in recovering the input parameters from the simulated data. Finally, in Section~\ref{sec:dc}, we summarize our conclusions and highlights in validating and refining reflection models to ensure their robustness in the era of the next-generation X-ray observatories.


\section{The ray-tracing code {\tt blackray}}\label{sec:ray_tra}

The shape of a relativistically broadened K$\alpha$ iron line is determined by several factors, including the background metric, the geometry of the emitting region, the disk emissivity, and the disk's inclination angle with respect to the line of sight of the distant observer. In the standard Kerr background framework, the spin parameter $a_\ast$ is the only relevant parameter of the background geometry, while the emitting region may range from the radius of the innermost stable circular orbit (ISCO) to some outer radius. For a specific geometry of the corona, the disk emissivity can be calculated with a ray-tracing code \citep[see, e.g.,][]{2022ApJ...925...51R}. For coronae with unknown geometry, it is common to employ a phenomenological emissivity profile like a power law, broken power law, or twice broken power law. In this study, for the sake of simplicity, we will employ a simple power law, where the emissivity of the disk is given by $\epsilon(r) \propto r^{-q}$, $r$ is the radial coordinate, and $q$ is the emissivity index.

Previous literature has discussed the computation of single iron lines in thin accretion disks \citep[see, e.g.,][]{Bambi2013,Bambi_2017}. In this study, we examine a distant observer with a viewing angle $i$ and we perform backward-in-time calculations of photon trajectories from the observer's image plane to the emission points on the accretion disk. The redshift factor $g = E_{\rm obs}/E_{\rm e}$, where $E_{\rm obs}$ is the photon energy at the detection point measured by the distant observer and $E_{\rm e}$ is the photon energy at the emission point measured in the rest-frame of the material in the disk, is calculated when a photon hits the disk using the formula
\begin{eqnarray}\label{eq:redshift}
g=\frac{ \sqrt{-g_{tt}-2g_{t\phi}\Omega - g_{\phi\phi} \Omega^2} }{ 1+b\Omega }, 
\end{eqnarray}
where $\Omega$ denotes the angular velocity of the fluid element on the disk at the emission point, $b = k_\phi/k_t$, and $k_t$ and $k_\phi$ denote the $t$ and $\phi$ components of the photon's 4-momentum. $b$ is a constant of motion, which can be determined from the photon's initial conditions. Finally, by integrating over the disk image, we obtain the iron line shape of the accretion disk using 
\begin{eqnarray}\label{eq:flux}
N(E_{\rm obs}) = \frac{1}{E_{\rm obs}} \frac{1}{D^2} \int g^3 I_{\rm e}(E_{\rm e}) \, dX \, dY , 
\end{eqnarray}
where $N(E_{\rm obs})$ is the photon count at the energy $E_{\rm obs}$ measured by the distant observer, $I_{\rm e}$ and $E_{\rm e}$ represent the specific intensity of radiation and the photon energy at the emission point on the disk (measured in the rest-frame of the gas in the disk), $X$ and $Y$ denote the Cartesian coordinates of the observer's image plane, and $D$ is the distance between the source and the distant observer. In the case of monochromatic emission with a power-law emissivity profile, the specific intensity of radiation $I_{\rm e}$ can be expressed as 
\begin{eqnarray}\label{eq:}
I_e(E_e)\propto\frac{\delta(E_{\rm e} - E_{\rm line})}{r^q} ,
\end{eqnarray}
where $E_{\rm line}$ is the energy of the emission line in the rest-frame of the material in the disk ($E_{\rm line} = 6.4$~keV in the case of the K$\alpha$ line of neutral iron, which will be used in the next section).

We utilize the ray-tracing code described in \citet{2019ApJ...878...91A}\footnote{The ray-tracing code can be downloaded from \url{https://github.com/ABHModels/blackray}.} to compute photon trajectories, but without calculating and tabulating the transfer function of the spacetime, as done in \citet{2019ApJ...878...91A}. The code is available on Zenodo~\citep{zenodo24}. With this approach, it is easier to have the control of the accuracy of the relativistic calculations, as we avoid the numerical uncertainties related to the calculation of the transfer function, its tabulation, and its integration. More specifically, the observer's screen is divided into a number of small elements (pixels), and the ray-tracing procedure gives each element's observed spectrum [which may be a redshifted/blueshift line if $I_{\rm e}$ is give by Eq.~(\ref{eq:}) or a redshifted/blueshift full reflection spectrum if $I_{\rm e}$ is extracted from the {\tt xillver} table as done in Subsection~\ref{ss-full}]. By summing up all of the elements, we obtain the total observed flux density of the disk. The accuracy of the calculations is regulated by the accuracy of the calculation of the photon trajectories, the size of the pixels, and the energy resolution of the spectrum. To sample the elements on the observer's screen, we use polar coordinates $(r_{scr}, \phi_{scr})$ that are related to the Cartesian coordinates $(X, Y)$ as $X=r_{scr}\cos{\phi_{scr}}$ and $Y=r_{scr}\sin{\phi_{scr}}$. The elements are incremented radially as $r_{scr}^{new}=1.0001\cdot r_{scr}^{current}$, and polarly as $\phi_{scr}^{new}=\phi_{scr}^{current}+0.1^\circ$. This sampling method enables us to obtain a detailed image of the entire accretion disk and capture all necessary and important information effectively. During the numerical integration of the photon trajectory, we require a precision tolerance in the range of $10^{-8}$ to $10^{-6}$ at each calculation step. To calculate the photon count per energy, $N(E_{\rm obs})$, a linearly spaced energy array with a binning size of 2.5~eV is used.


\section{Simulations}\label{sec:simu}

\subsection{Current reflection models}
\label{current}

\begin{figure*}
\includegraphics[width=0.48\linewidth]{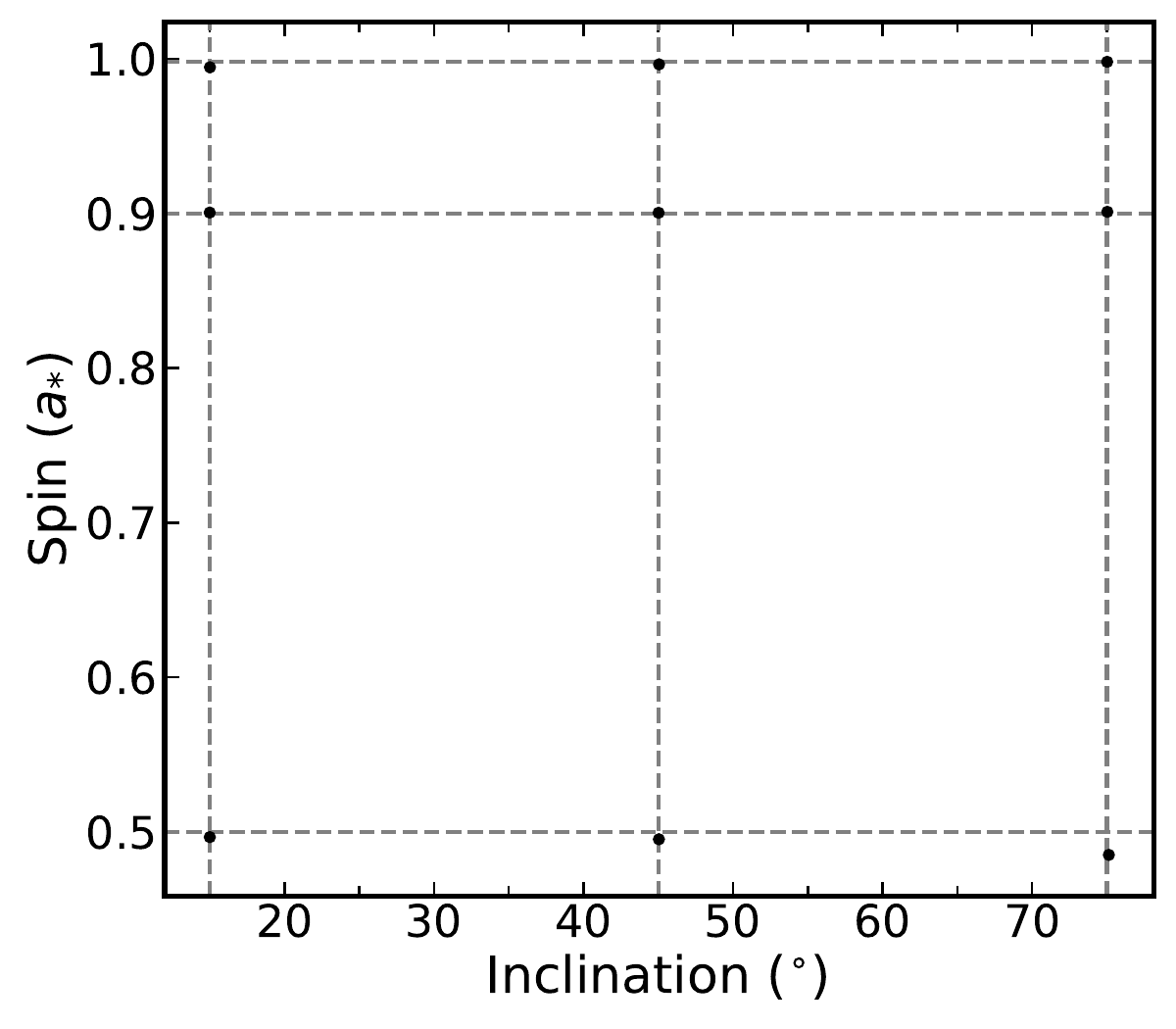}
\includegraphics[width=0.48\linewidth]{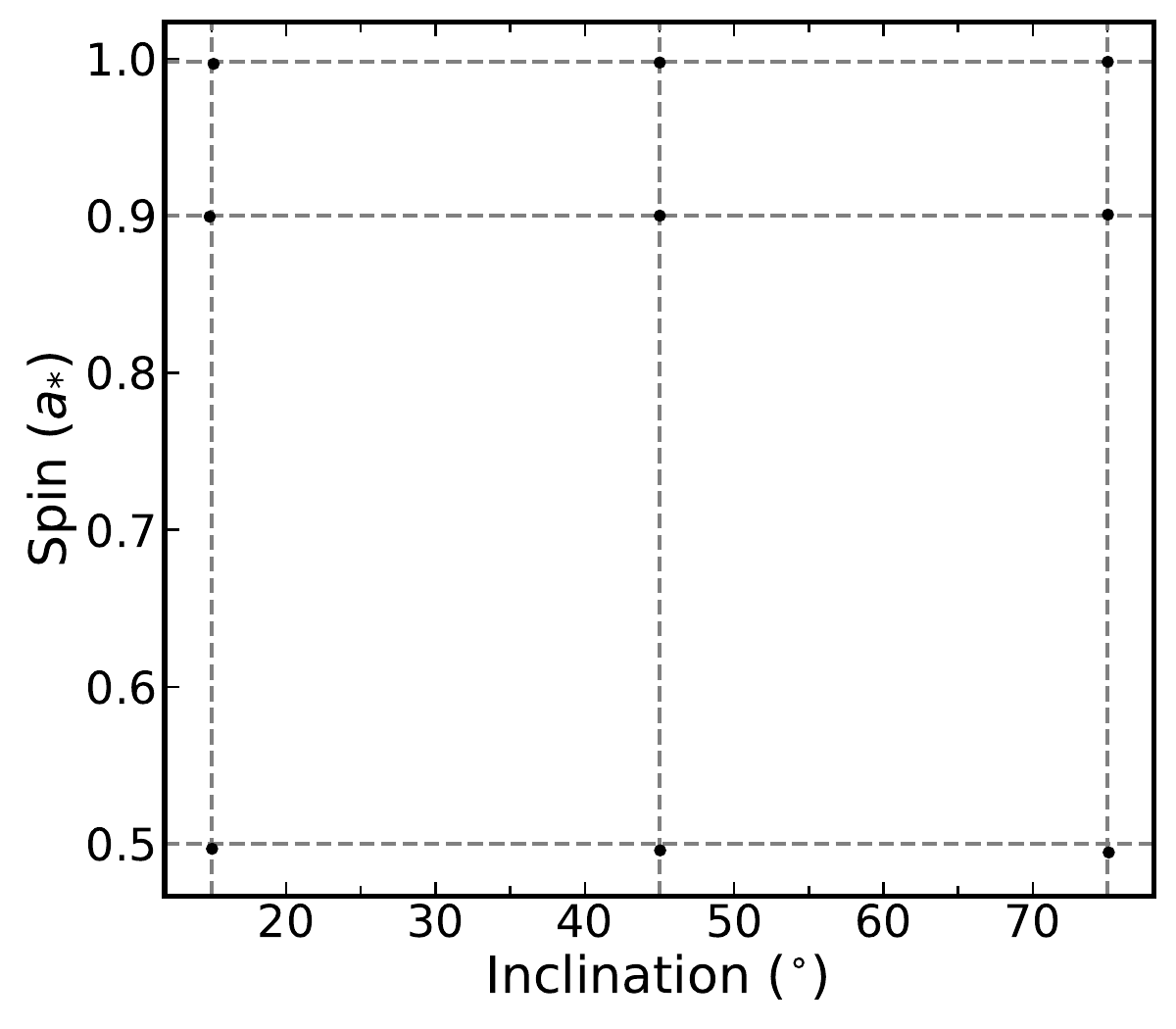}
\caption{Constraints on the spin parameter and inclination angle after fitting the simulated spectra with the current version of {\tt relline}. The crosses of dashed grey lines represent the input values for the simulations and the black dots indicate the best-fit values of the two parameters. The error bars (90\% CL) are smaller than the size of the dots. The left panel is for the case with emissivity index $q=3$ and the right panel is for $q=5$.}
\label{fig:constraint_relline}
\end{figure*}

To see if the current reflection models are accurate enough for the high quality data promised by future instruments, we simulate a set of observations with the X-ray Integral Field Unit (X-IFU) onboard \textsl{Athena} \citep{2013arXiv1306.2307N}\footnote{We use the latest response files of \textsl{Athena} as of April 2024: \url{https://x-ifu.irap.omp.eu/resources/for-the-community}.} and the Large Area Detector (LAD), which may also fly in future. The simulations are undertaken with the {\tt fakeit} command in {\tt XSPEC} v12.11.1 \citep{1996ASPC..101...17A}. We assume an exposure time of 100~ks for each instrument. The input models for the simulations are the iron lines calculated with the ray-tracing code described in Section~\ref{sec:ray_tra} plus a simple power law with the photon index $\Gamma=1.6$ to describe the emission from the corona. We set the source flux to $1\times10^{-7}$~erg~s$^{-1}$~cm$^{-2}$ in the 2--10~keV band\footnote{This can be the flux of an exceptionally bright Galactic black hole, like MAXI~J1820+070 in its 2018 outburst \citep[e.g.][]{Shidatsu2019ApJ...874..183S} or 4U~1543--47 in its 2021 outburst \citep[e.g.][]{Connors2021ATel14725....1C}.} and an equivalent width (EW) of 250~eV for the iron line. A total number of 18 spectra are simulated on the grids of black hole spins $a_*=0.5$, 0.9, and 0.998, inclination angles $i=15$, 45, and 75~deg, and indices for the emissivity profile $q=3$ and 5 (see Fig.~\ref{fig:constraint_relline}). The inner disk radius is fixed at the ISCO and the outer radius is fixed at 500~$R_{\rm g}$ ($R_{\rm g}=GM/c^2$ is the gravitational radius). The spectra are then grouped using the ``optimal binning'' strategy \citep{Kaastra2016A&A...587A.151K} with the tool {\tt ftgrouppha}\footnote{\url{https://heasarc.gsfc.nasa.gov/lheasoft/ftools/headas/ftgrouppha.html}}.

\begin{figure*}
\includegraphics[width=0.48\linewidth]{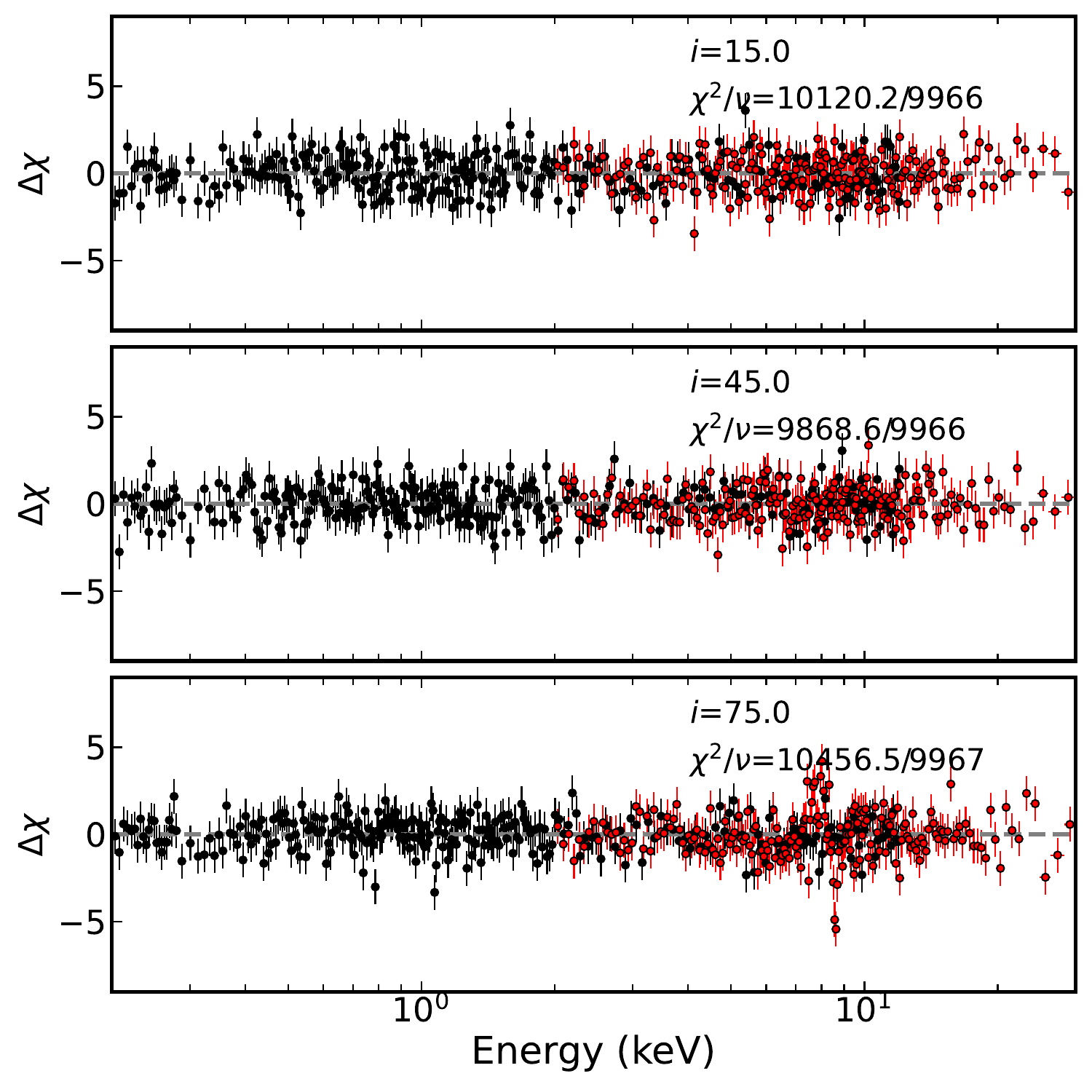}
\includegraphics[width=0.48\linewidth]{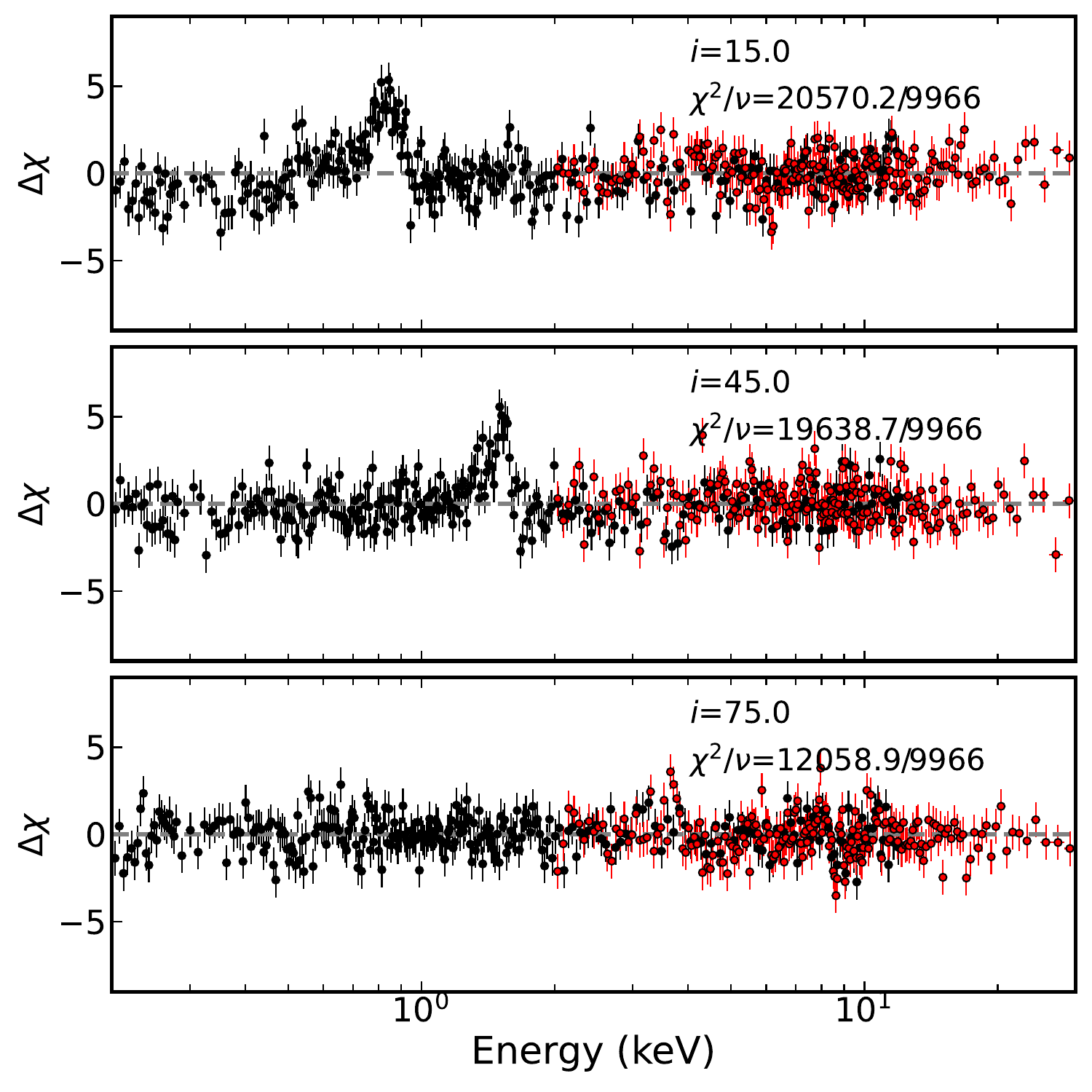}
\caption{Residuals of fitting the simulated spectra with the model {\tt relline}. Only those with $a_*=0.998$ are shown. The left panel represents the case when the emissivity index is $q=3$ and the right is for $q=5$. The black color represents data of \textsl{Athena}/X-IFU and red is for LAD.}
\label{fig:residual}
\end{figure*}

We fit the spectra with the model {\tt powerlaw+relline} where {\tt relline} \citep{Dauser2010MNRAS.409.1534D} is a module from the package {\tt relxill}~v2.0 \citep{2014ApJ...782...76G}. It is to fit the relativistic broadening of a single line (6.4~keV in this case) from the accretion disk. The constraints on the spin and inclination parameters are shown in Fig.~\ref{fig:constraint_relline}. We can see that the parameters are recovered well, with the best-fit values close to the input and the error bars too small to be seen. Fig.~\ref{fig:residual} shows the residuals of the best-fit models for these selected simulations.

\begin{figure}
\centering
\includegraphics[width=0.99\linewidth]{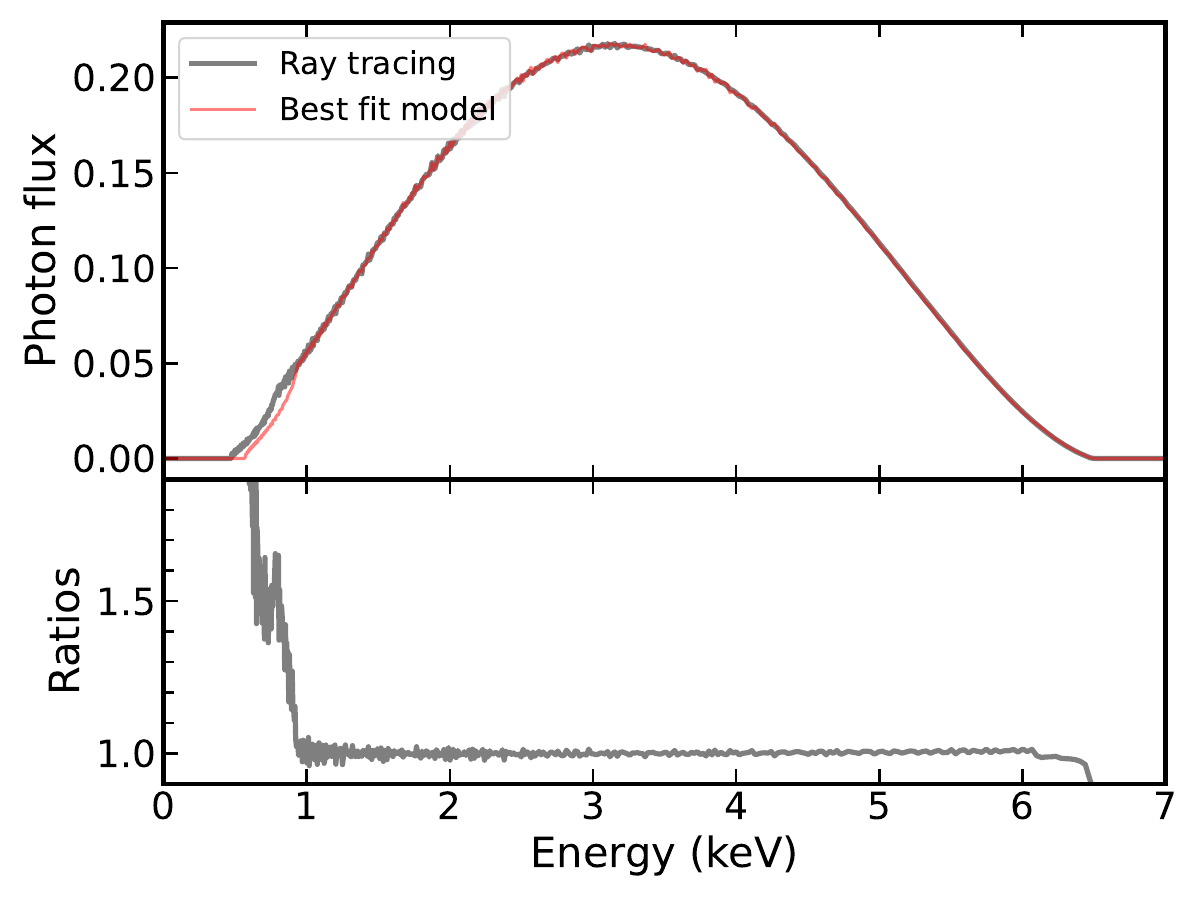}
\caption{Iron line profile generated with the ray-tracing code and best-fit model of {\tt relline} (top panel) and their ratio plot (bottom panel) for the case $a_*=0.998$, $i = 15$~deg, and $q=5$.} 
\label{fig:residualexplanation}
\end{figure}

In the case of $q=3$, the two models fit the data well and there are no significant unresolved features. However, large discrepancy are present for $q=5$, which appears to be more significant when the inclination angle gets lower. The line feature in the residuals of the simulations with $q=5$ and $i = 15$ and 45~deg is produced by a discrepancy in the low energy tail of the iron line between the ray-tracing code and {\tt relline} (see Fig.~\ref{fig:residualexplanation} for the case $i = 15$~deg). Fittings with \texttt{relline\_nk} follow the same pattern. It indicates that current reflection models may not be accurate enough for high quality data expected from \textsl{Athena}/X-IFU and LAD. This raises concerns about reflection analysis on data from future X-ray missions since a steep emissivity ($q>3$) has been commonly found in X-ray binaries \citep[e.g.][]{Liu2022arXiv221109543L,Liu2022MNRAS.513.4308L} and active galactic nuclei \citep[e.g.][]{Jiang2018MNRAS.477.3711J}. More accurate models would be required for these cases.

\begin{figure*}
\includegraphics[width=0.48\linewidth]{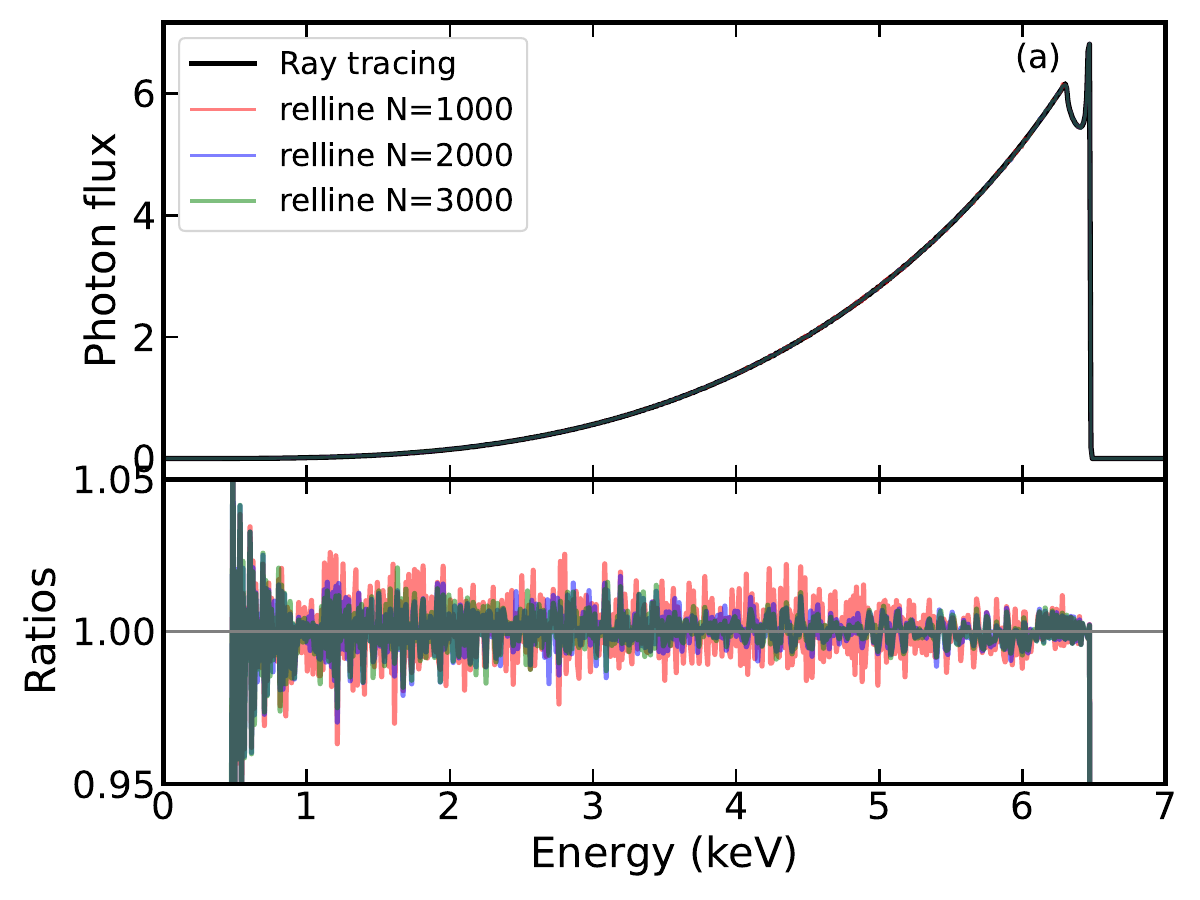}
\includegraphics[width=0.48\linewidth]{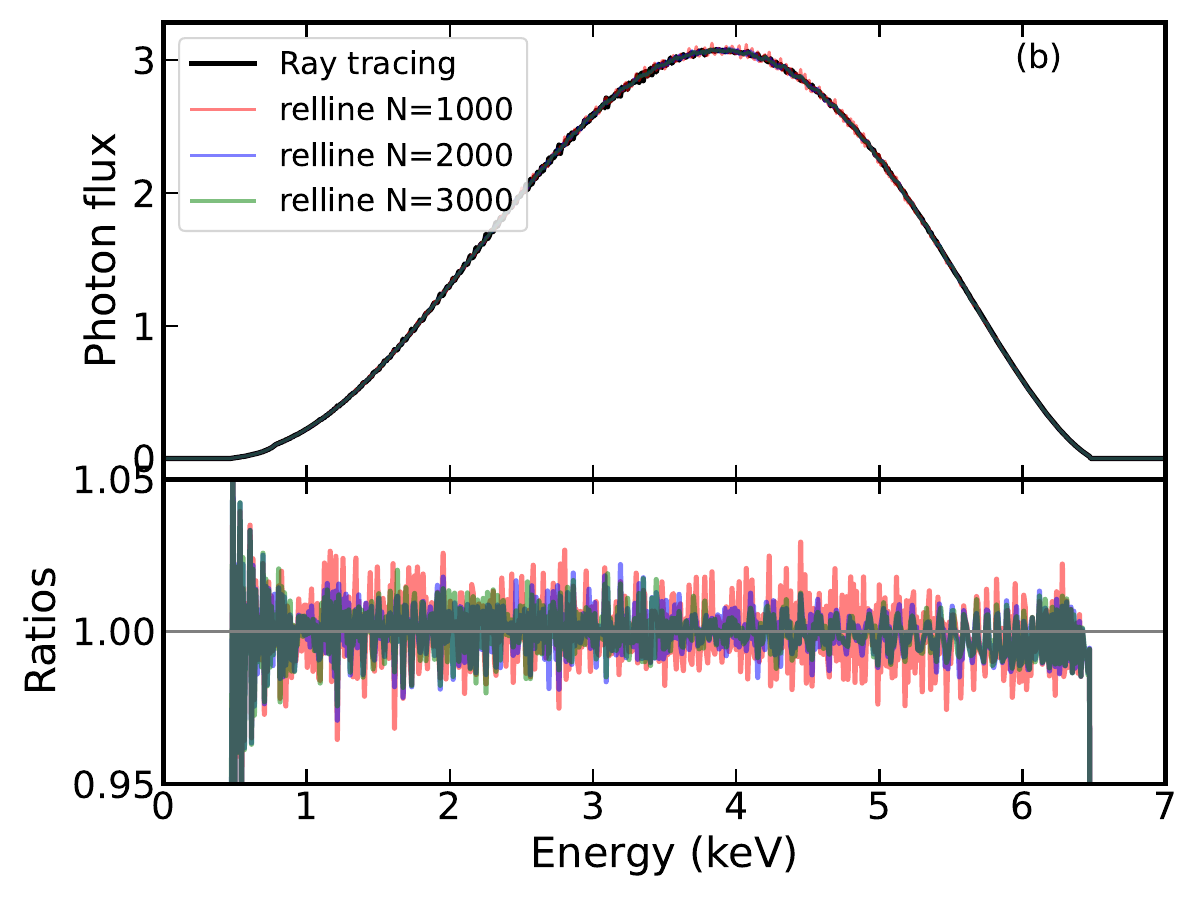} \\
\includegraphics[width=0.48\linewidth]{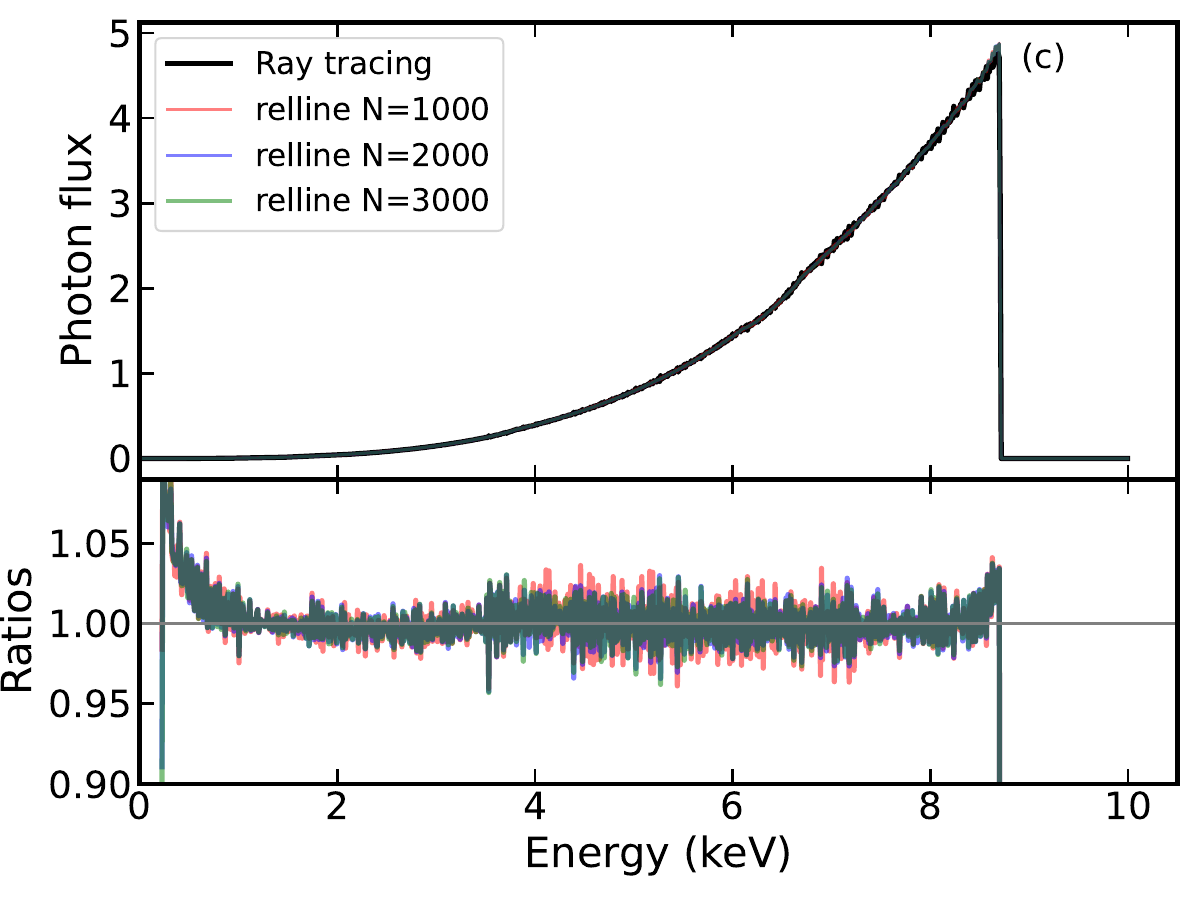}
\includegraphics[width=0.48\linewidth]{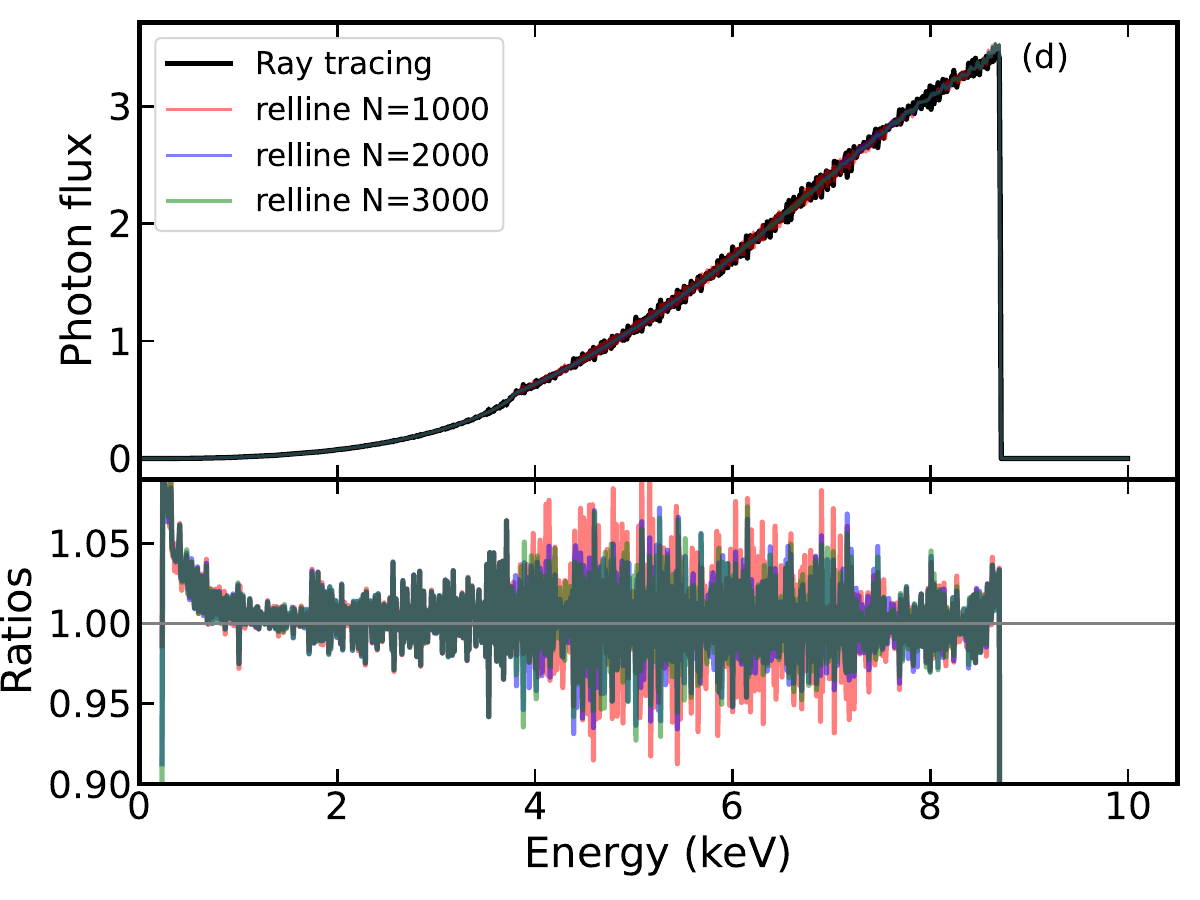}
\caption{Difference between the broad iron line from the ray-traing code and \texttt{relline} model for N\_FRAD = 1000, 2000, and 3000. Panel (a) for $a_*=0.998$, $i=15$ and $q=3$, (b) for $a_*=0.998$, $i=15$ and $q=5$, (c) for $a_*=0.998$, $i=75$ and $q=3$ and (d) for $a_*=0.998$, $i=75$ and $q=5$.}
\label{fig:relline_diff}
\end{figure*} 

\begin{figure*}
    \includegraphics[width=0.48\linewidth]{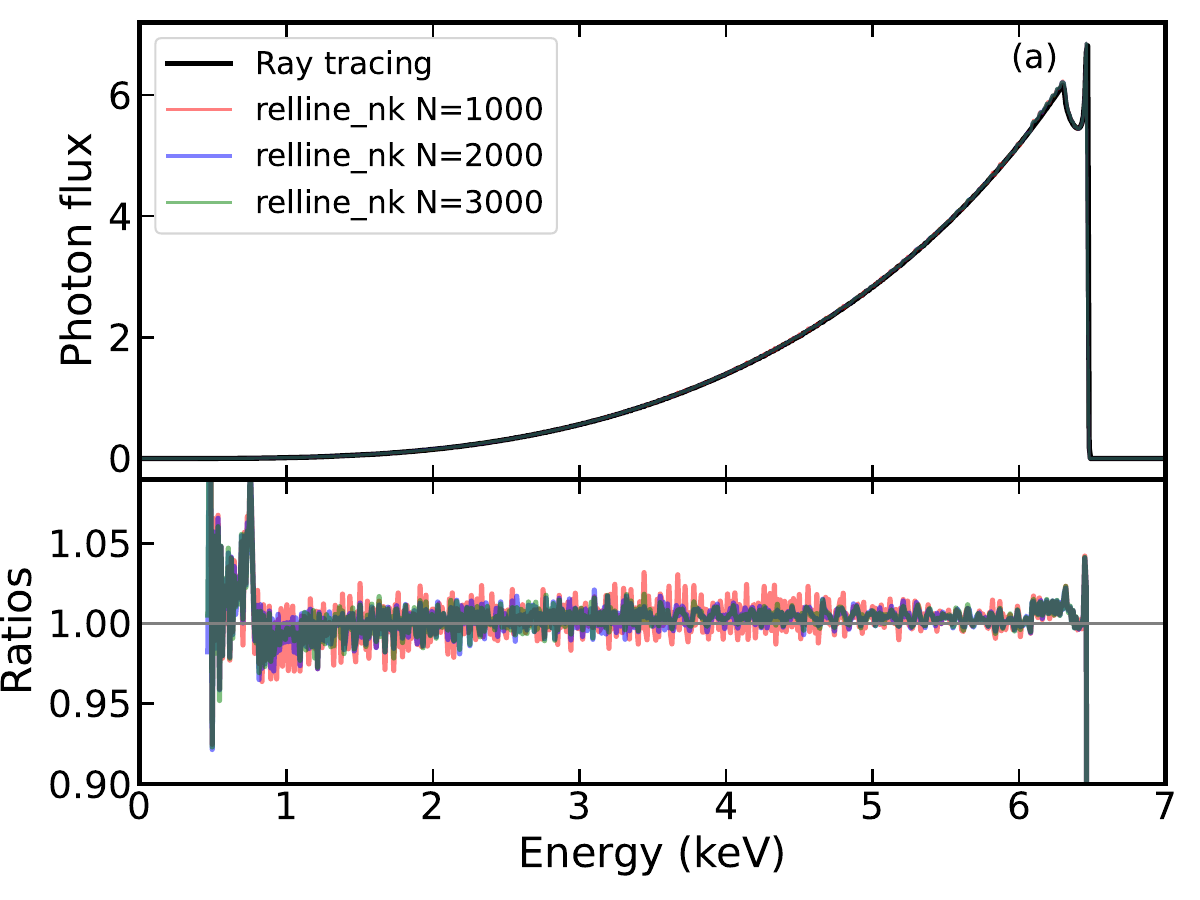}
    \includegraphics[width=0.48\linewidth]{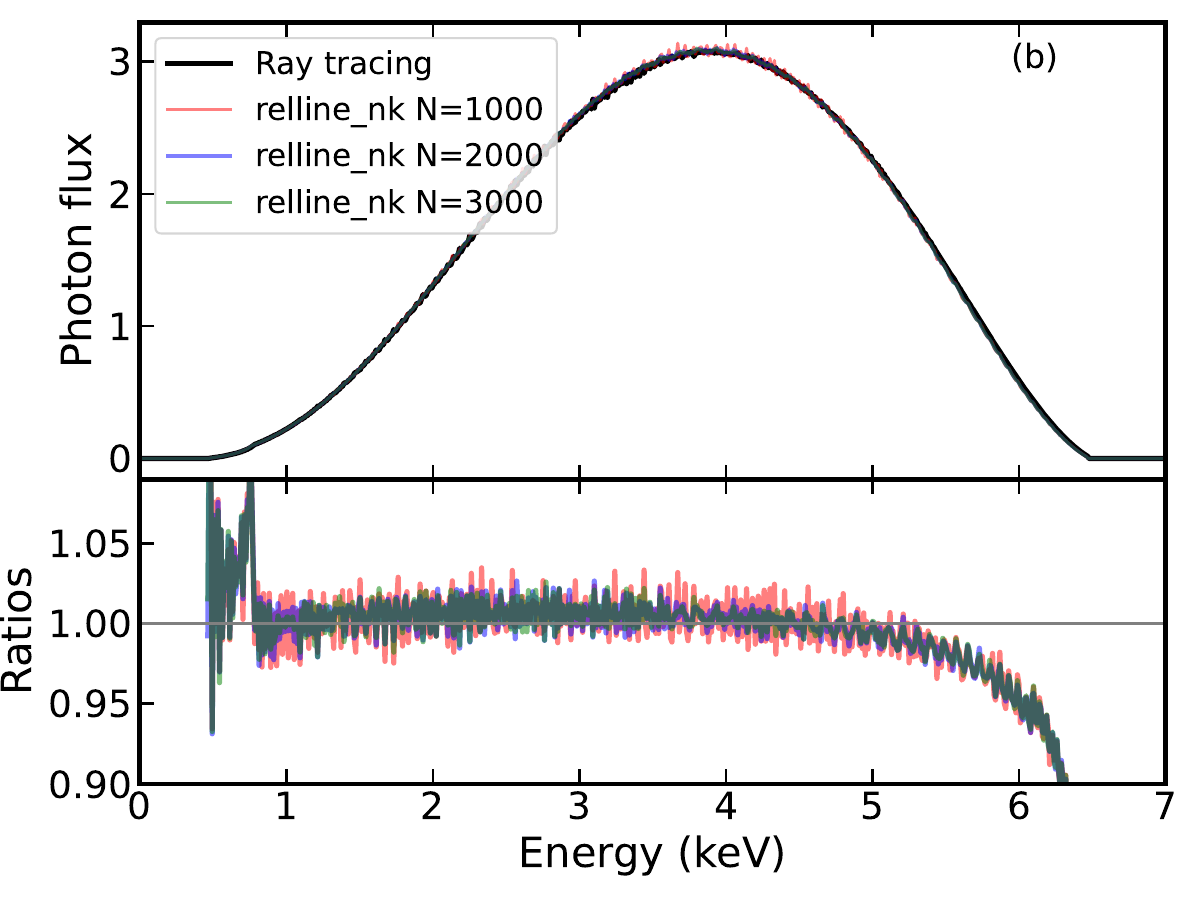} \\
    \includegraphics[width=0.48\linewidth]{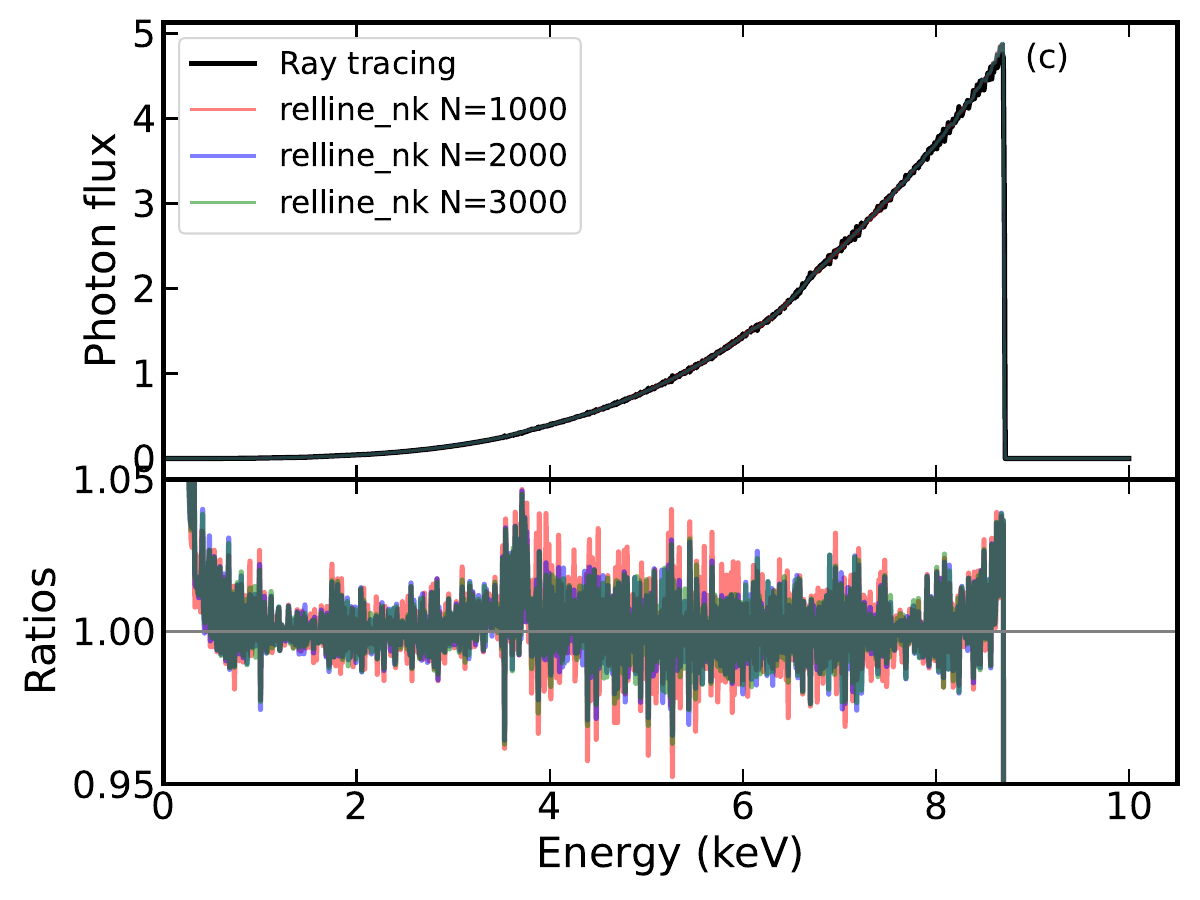}
    \includegraphics[width=0.48\linewidth]{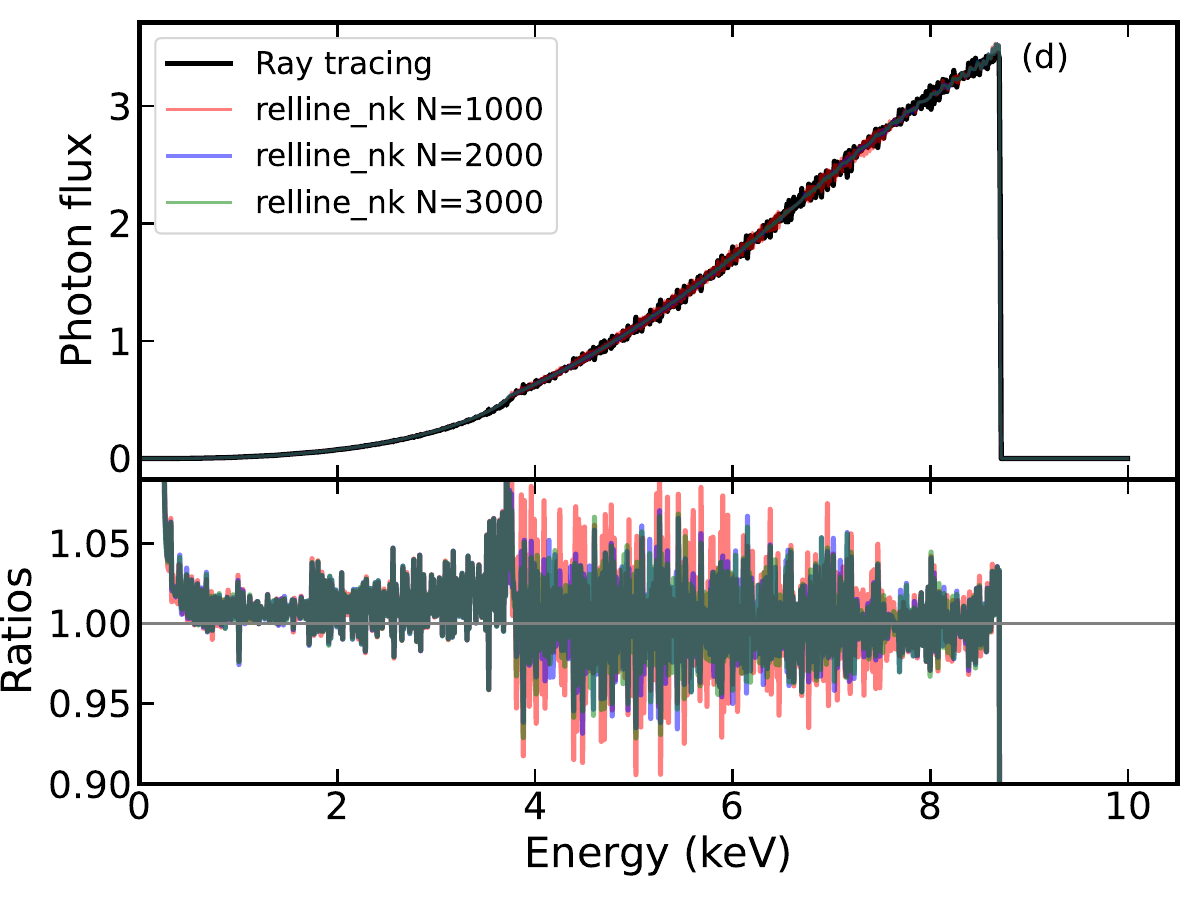}
\caption{The same as Fig.~\ref{fig:relline_diff} but for model \texttt{relline\_nk}.}
\label{fig:relline_nk_diff}
\end{figure*}

\subsection{More accurate reflection models}
\label{new_model}

Ray-tracing calculations are too time consuming to be done during the data analysis process. For this reason, {\tt relxill} and {\tt relxill\_nk} employ the formalism of the transfer function proposed by Cunningham~\citep{1975ApJ...202..788C}. The integral in Eq.~(\ref{eq:flux}) is rewritten as
\begin{eqnarray}
\label{eq-Fobs}
N(E_{\rm obs}) = \frac{1}{E_{\rm obs}} 
 \frac{1}{D^2} \int_{r_{\rm in}}^{r_{\rm out}} \int_0^1
\pi r_{\rm e} \frac{ g^2 \, f \, I_{\rm e}}{\sqrt{g^* (1 - g^*)}} \, dg^* \, dr_{\rm e} \, , \nonumber\\
\end{eqnarray}
where the integration is now on the accretion disk ($r_{\rm in}$ and $r_{\rm out}$ are, respectively, the inner and the outer edge of the accretion disk, while $r_{\rm e}$ is the emission radius) rather than on the image plane of the distant observer and we have introduced the transfer function $f = f(g^*,r_{\rm e},i)$:
\begin{eqnarray}\label{eq-trf}
f(g^*,r_{\rm e},i) = \frac{1}{\pi r_{\rm e}} g 
\sqrt{g^* (1 - g^*)} \left| \frac{\partial \left(X,Y\right)}{\partial \left(g^*,r_{\rm e}\right)} \right| \, .
\end{eqnarray}
$g^*$ is the relative redshift factor
\begin{eqnarray}
g^* = \frac{g - g_{\rm min}}{g_{\rm max} - g_{\rm min}} \, ,
\end{eqnarray}
which ranges from 0 to 1. $g_{\rm max}=g_{\rm max}(r_{\rm e},i)$ and $g_{\rm min}=g_{\rm min}(r_{\rm e},i)$ are, respectively, the maximum and the minimum values of the redshift factor $g$ for the photons emitted from the radial coordinate $r_{\rm e}$ and detected by a distant observer with polar coordinate $i$. The quantity $\left| \partial \left(X,Y\right)/\partial \left(g^*,r_{\rm e}\right) \right|$ appearing in the expression of the transfer function is the Jacobian. In {\tt relxill} and {\tt relxill\_nk}, transfer functions for a grid of spacetimes and disk's inclination angles are pre-calculated and tabulated in FITS files. During the data analysis process, the models call the FITS file and can quickly calculate the integral in Eq.~(\ref{eq-Fobs}).

Eq.~(\ref{eq-Fobs}) directly follows from Eq.~(\ref{eq:flux}) after changing the integration variables. Systematic uncertainties in the models for data analysis are related to the fact we use a grid of transfer functions (so the models have to calculate a transfer function for a generic spacetime and disk's inclination angle through interpolation) and every transfer function is tabulated for a finite number of radii. {\tt relxill} and {\tt relxill\_nk} also assume that $I_{\rm e}$ is the same throughout the disk (but this will not affect our comparison with the predictions of the ray-tracing code because we impose the same assumption when we calculate the spectra with the ray-tracing code). In their latest versions, {\tt relxill} has a FITS file with a grid 25~spin parameters $\times$ 30~inclination angles, while {\tt relxill\_nk} employs a FITS file with a grid 30~spin parameters $\times$ 30~deformation parameters $\times$ 22~inclination angles\footnote{We note that {\tt reltrans}~\citep{2019MNRAS.488..324I} does not use a FITS file grid of transfer functions and instead it calculates the reflection kernel on the fly. It uses by default a 300 by 300 oval grid in $X$ and $Y$ and presumably the accuracy of the relativistic calculations is very similar to that of {\tt relxill} and {\tt relxill\_nk}.}. Every point of the grid of the FITS file represents a specific spacetime observed from a certain viewing angle and has its own transfer function, so we have in total 750~transfer functions in the FITS file of {\tt relxill} and 19800~transfer functions in the FITS file of {\tt relxill\_nk}. Every transfer function is tabulated for 100~radii (from the ISCO to $\sim 1000$~$R_{\rm g}$) and 40~values of the relative redshift $g^*$. We checked if the residuals in the fits could be removed increasing the number of grid points in the FITS files or the number of radii or relative redshifts in every transfer functions, but no significant difference was found.

\begin{figure*}
\includegraphics[width=0.45\linewidth]{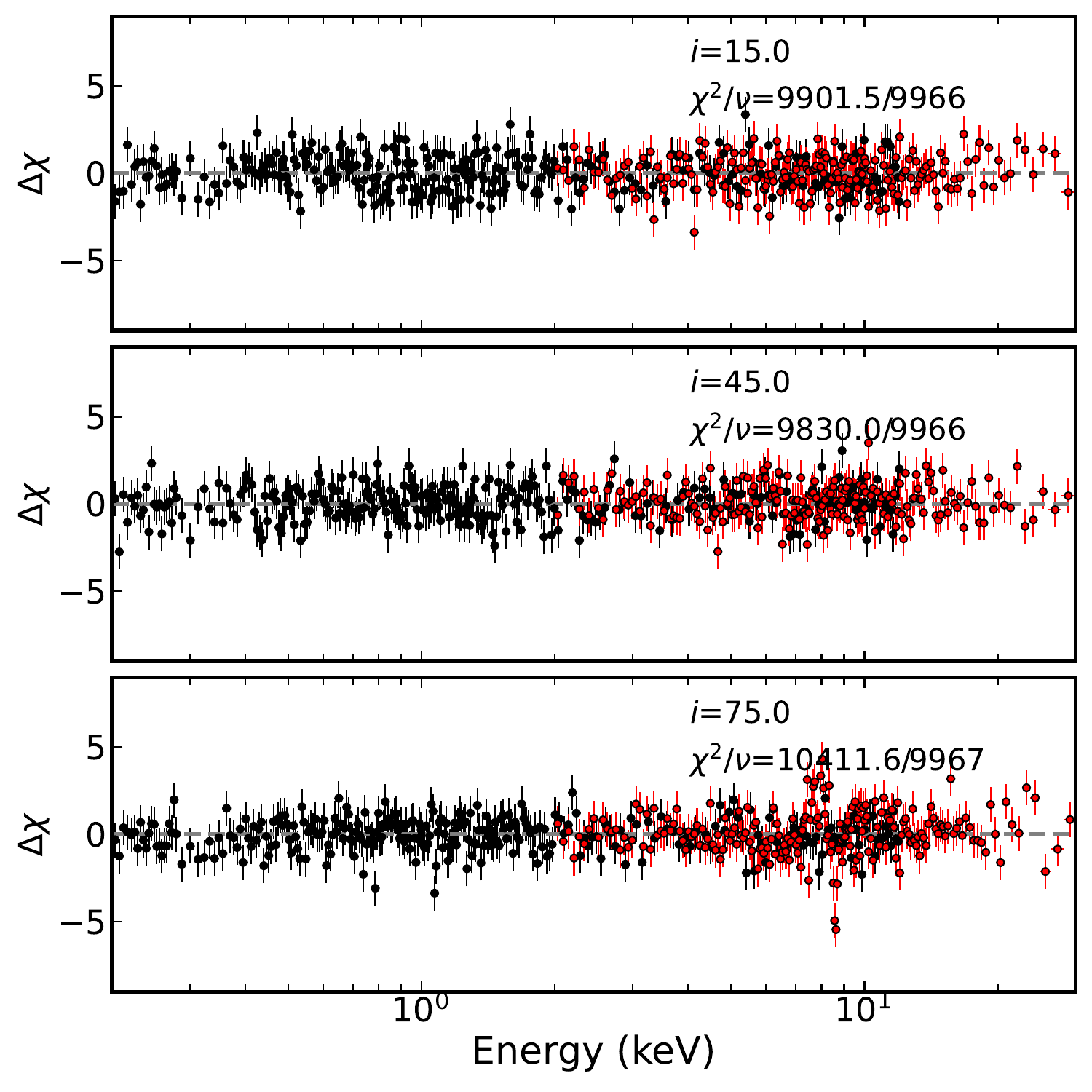}
\includegraphics[width=0.45\linewidth]{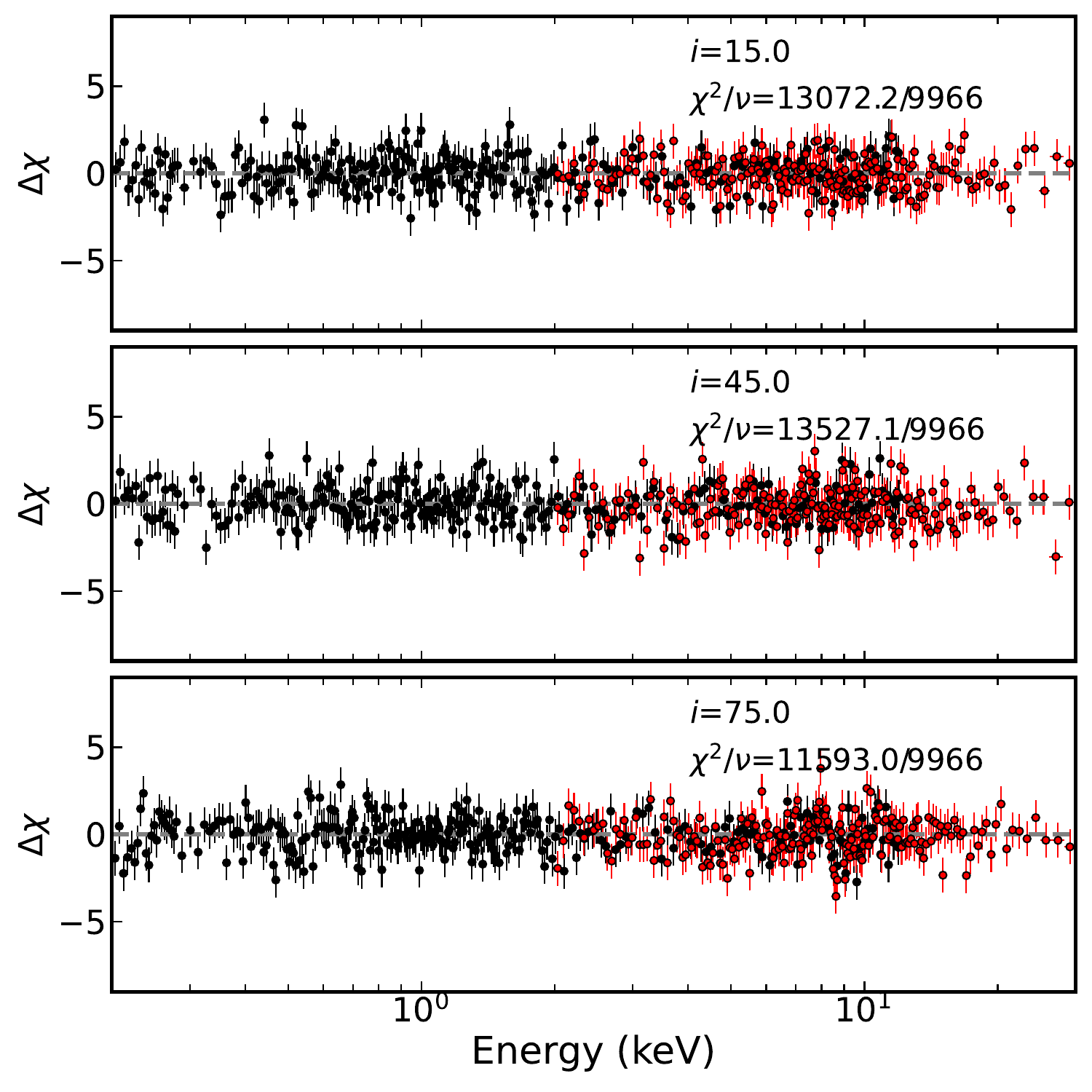}
\caption{Residuals of fitting the simulated spectra for $a_*=0.998$ with the {\tt relline} model in which the number of interpolation points on the disk (N\_FRAD) is set to 3000. See the text in Section~\ref{new_model} for more details. The left panel represents the case $q=3$ and the right is for $q=5$. Colors are coded as in Fig.~\ref{fig:residual}}. 
\label{fig:residual_new}
\end{figure*}

We found that enhancing the number of emitting points on the accretion disk utilized in the calculations of the resulting flux through interpolations within the models effectively solves the problem. Consequently, we modified the variable N\_FRAD of the models from 1000 to 3000, without changing the number of points for $g^*$. With such an improvement in the calculation of the integral, we attain a more precise line broadening model without modifying the FITS file for the transfer function. In Fig.~\ref{fig:relline_diff}  (Fig.~\ref{fig:relline_nk_diff}), we show the difference between the ray-traced iron lines and \texttt{relline} (\texttt{relline\_nk}). Increasing the N\_FRAD parameter apparently improves the precision of \texttt{relline} (\texttt{relline\_nk}). Applying this model to the simulated spectra with $a_*=0.998$ and $q=5$ provides good fits without significant residuals (see Fig.~\ref{fig:residual_new}).

\subsection{Non-Kerr spacetime}
\label{non-kerr}

\begin{figure}
    \includegraphics[width=0.99\linewidth]{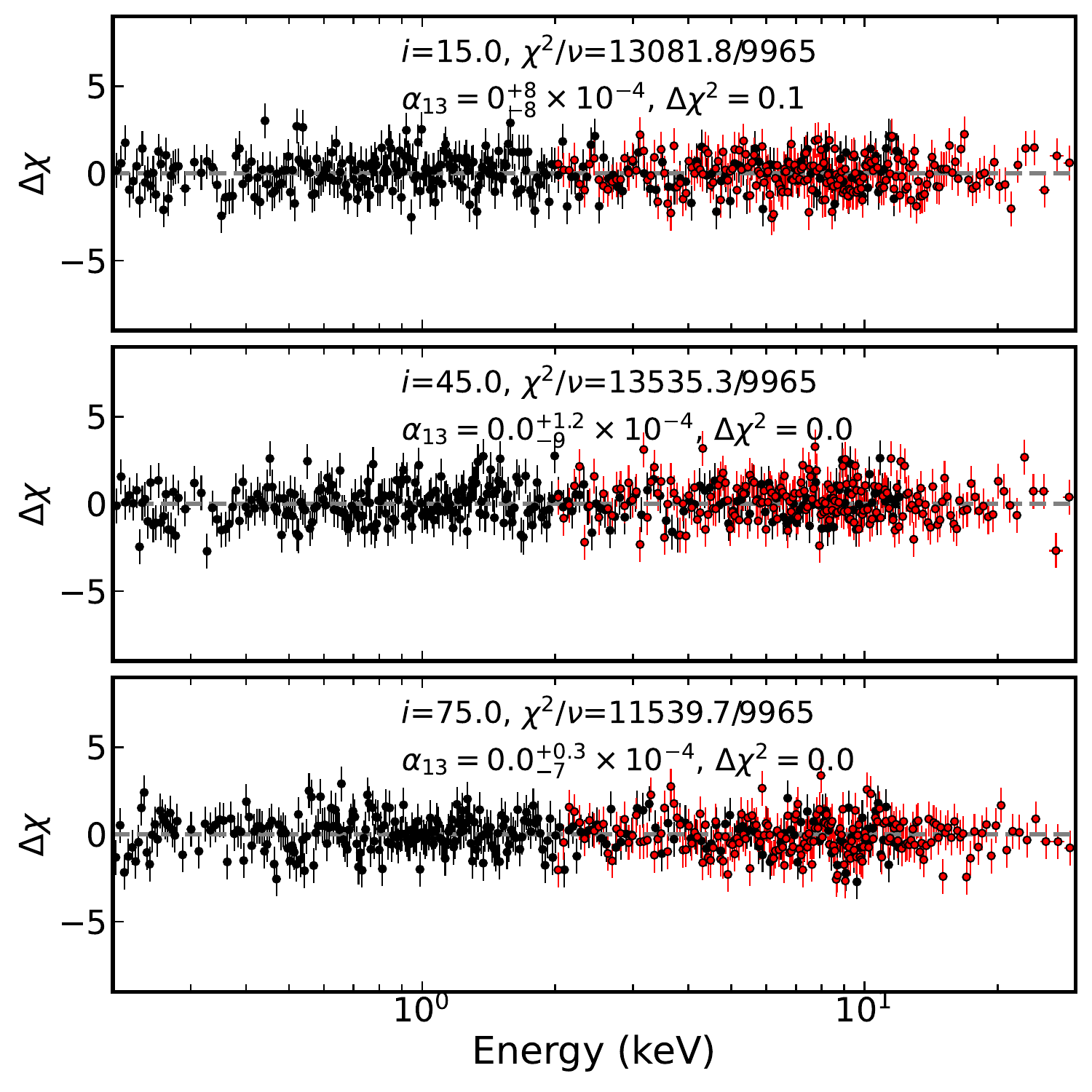}
    \caption{Residuals of fitting the simulated spectra for $a_*=0.998, q=5$ with the {\tt relline\_nk} model in which the number of interpolation points on the disk (N\_FRAD) is set to 3000. The $\Delta\chi^2$ values represent the improvement in $\chi^2$ compared to the case of $\alpha_{\rm 13}=0$. See the text in Section~\ref{non-kerr} for more details.}. 
    \label{fig:residual_a13_free}
\end{figure}

Until now, we have been testing the reflection models under the framework of the Kerr spacetime as predicted by General Relativity. In this section, we evaluate the capability of the more accurate reflection model in testing General Relativity. We fit the spectra in the right panels of Fig.~\ref{fig:residual_new} with the model {\tt powerlaw + relline\_nk} where {\tt relline\_nk} includes a deformation parameter $\alpha_{13}$ to quantify possible deviations from Kerr spacetime \citep[see][for details]{Bambi_2017}. Here the spacetime geometry is described by the Johannsen metric \citep{2013PhRvD..88d4002J}, which exactly reduces to the Kerr solution when $\alpha_{13} = 0$ while it deviates from the Kerr solution for any non-vanishing value of the deformation parameter $\alpha_{13}$. The residuals, statistics and constraints on the deformation parameter for this set of fits are shown in Fig.~\ref{fig:residual_a13_free}. At the 90\% confidence level, we always recover the Kerr solution.

\subsection{Simulation with full reflection models}
\label{ss-full}

In this section, we test the accuracy of current reflection models using full reflection spectra, instead of a simple broad iron line. We first generate a set of full reflection spectra using the ray-tracing code. We consider two cases: (1) the emission angle ($\theta_{\rm e}$) from the disk surface is equal to the inclination angle ($i$) of the observer. (2) the correct emission angles are chosen for each site on the disk when calling the \texttt{xillver} table.

\begin{figure*}
\includegraphics[width=0.45\linewidth]{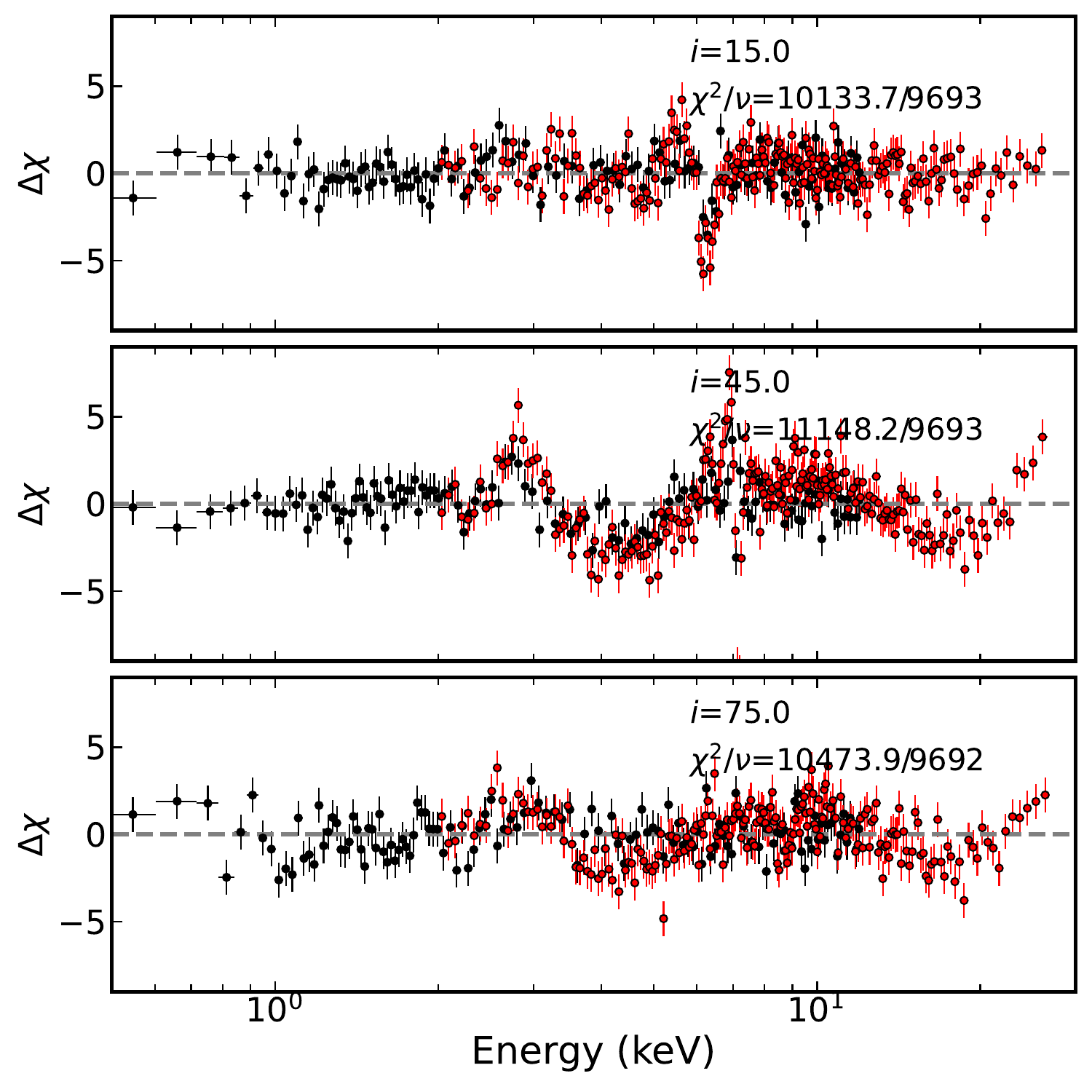}
\includegraphics[width=0.45\linewidth]{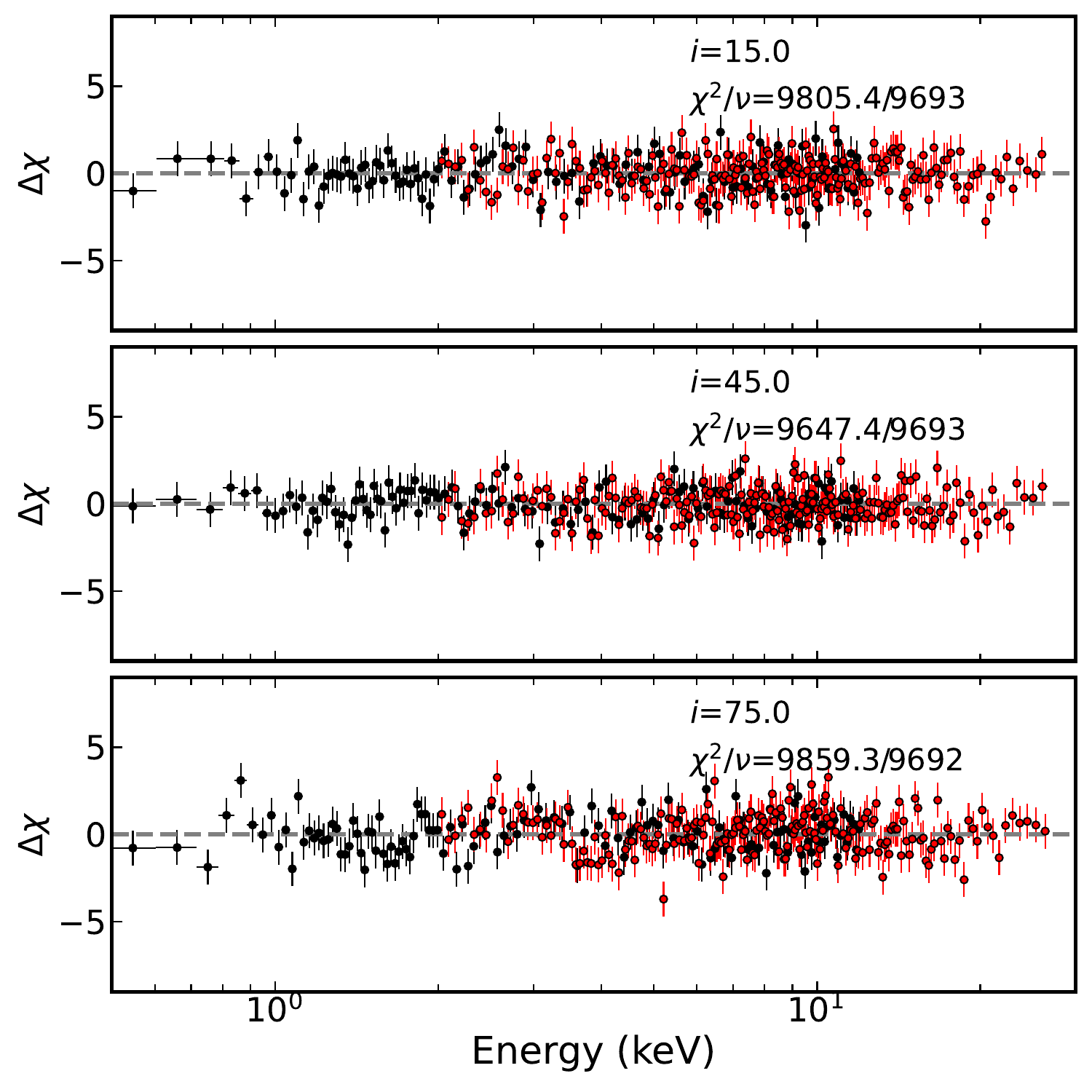}
\caption{Residuals of fitting the simulated full reflection spectra for $a_*=0.5$ and $q=5$, which is the case showing the largest residuals in our set of simulations. The left panel represents fittings with current models and the right is for when increasing the parameter N\_ENER\_CONV from 4096 to 524288. Colors are coded as in Fig.~\ref{fig:residual}}. 
\label{fig:residual_relconv}
\end{figure*}

Case I -- The ray-traced full reflection spectra should be identical to \texttt{relconv*xillver} in \texttt{XSPEC}. We generate 18 full reflection spectra with spins $a_*=0.5$, 0.9, and 0.998, inclination angles $i=15$, 45, and 75~deg, and power-law emissivity profile with the indices $q=3$ and 5. The other parameters required for the calculation are the photon index ($\Gamma=1.7$), ionization parameter ($\log\xi=1.0$), cutoff energy ($E_{\rm cut}=300$~keV), and iron abundance ($A_{\rm Fe} = 1$, namely Solar iron abundance). To mimic real observations, we also include a Galactic absorption component (\texttt{tbabs}) with the column density set to $N_{\rm H}=0.6\times 10^{22}$~cm$^{-2}$. As in Sec.~\ref{current}, we set a flux of $1\times10^{-7}$~erg~s$^{-1}$~cm$^{-2}$ in the 2--10~keV band. Additionally, we set equal flux for the reflection component and the \texttt{cutoffpl} component in the 20--40~keV band. These settings represent observations of bright black hole X-ray binaries in the hard state. The simulated spectra are fitted with the model \texttt{const*tbabs*(cutoffpl+relconv*xillver)} in \texttt{XSPEC}. In some cases, the data can be fitted well but significant residuals are seen in other cases (see the left panels of Fig.~\ref{fig:residual_relconv}). The residuals cannot be removed by increasing the N\_FRAD parameter as in the simulations for iron lines, indicating that it is not related to the accuracy of the transfer function. We find that the residuals are removed if we increase another parameter, N\_ENER\_CONV, which regulates the number of energy bins and thus the energy resolution in the reflection kernel, from 4096 to 524288 (see the right panels of Fig.~\ref{fig:residual_relconv}).

\begin{figure}
    \includegraphics[width=0.99\linewidth]{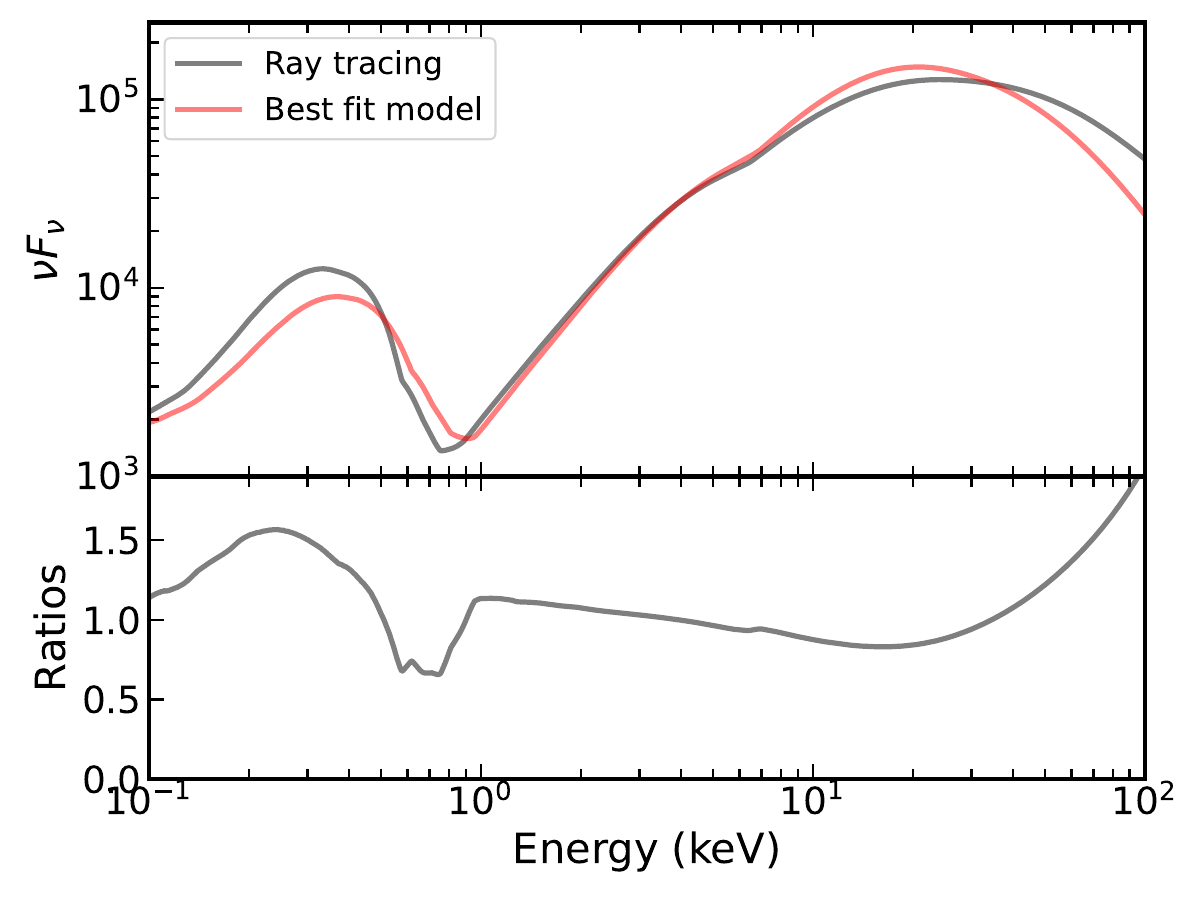}
    \caption{Example of a full reflection spectrum generated with the ray-tracing code and best-fit model of {\tt relxill} (top panel) and their ratio plot (bottom panel) in Case~II. In this simulation, we assume $a_* = 0.998$, $i = 15$~deg, $\Gamma = 1.7$, $E_{\rm cut} = 300$~keV, $q = 5$, $\log\xi = 1.0$, and $A_{\rm Fe} = 1$.} 
    \label{fig:diffff}
\end{figure}

Case II -- The second case represents the most accurate calculations of full reflection spectra from the accretion disk. We simulate 18 spectra using the same procedure as above and fit the spectra with the model \texttt{const*tbabs*(cutoffpl+relxill)}. Based on previous simulations with the iron line and Case I, we set N\_ENER\_CONV=524288 and N\_FRAD=3000 for \texttt{relxill}. However, no acceptable fittings can be found, as $\chi^2/\nu>50$ for all fittings. Fig.~\ref{fig:diffff} shows the spectrum predicted with the ray-tracing code and the best-fit model of {\tt relxill} for the case $a_* = 0.998$, $i = 15$~deg, $\Gamma = 1.7$, $E_{\rm cut} = 300$~keV, $q = 5$, $\log\xi = 1.0$, and $A_{\rm Fe} = 1$. The reason for these residuals is that \texttt{relxill} uses some simplifications, averaging the emission angle in every radial zone instead of using the actual \texttt{xillver} spectra for the emission angle in the convolutional process. This approach is taken to reduce the computational time of the convolution in the kernel. In an attempt to further improve the full reflection model, we implement in \texttt{relxill\_nk} 50~radial zones, compared to the one-zone approximation in \texttt{relxill}. This makes \texttt{relxill\_nk} more accurate than \texttt{relxill}. However, we find that the new model does not necessarily improve the fitting statistics.

\section{Discussion and conclusions}\label{sec:dc}

For our simulations, we have considered an exceptionally bright Galactic black hole with a flux of $1\times10^{-7}$~erg~s$^{-1}$~cm$^{-2}$ in the 2--10~keV band. We considered the case in which the input model is simply a relativistically broadened iron line and a power law. These choices can maximize any minor discrepancy in the relativistic calculations of {\tt relxill} and {\tt relxill\_nk}. For a more conventional Galactic black hole with a lower flux, the statistical uncertainty in the photon count increases and minor residuals may disappear. Even the choice of considering only an iron line is to identify better possible discrepancies between the ray-tracing code calculations and the the reflection models {\tt relxill} and {\tt relxill\_nk}.

Assuming the Kerr metric (Section~\ref{current}), we find that we always recover the correct input parameters, but the fit shows some residuals in the case of higher emissivity index ($q=5$). We note that we obtain qualitatively similar results from {\tt relline} and {\tt relline\_nk}, suggesting that the problem is not in the accuracy of the calculations of the transfer functions but in how the transfer function is tabulated or how it is integrated to infer the observed spectrum. The fact that we do not see residuals for the case $q=3$ and we have instead residuals when $q=5$ suggests that the problem is in the accuracy of the calculations at small radii, near the black hole.

As discussed in the previous section, such a discrepancy is found to be generated by the accuracy of the integration of the transfer function in the reflection models {\tt relxill} and {\tt relxill\_nk}. In the current versions, the number of interpolation points on the disk is N\_FRAD = 1000. If we set N\_FRAD = 3000, we significantly improve the fit and the residuals disappear (Section~\ref{new_model}).

In conclusion, with simulations based on iron lines, we have shown that the accuracy of the relativistic calculations in the current versions of {\tt relxill} and {\tt relxill\_nk} (which are encoded in the transfer functions and tabulated in the corresponding FITS files) is good enough for the next generation of X-ray detectors like X-IFU and LAD, even in the case of an exceptionally bright Galactic black hole. The two reflection models require a minor modification in the subroutine integrating the transfer function to obtain the final spectrum. Such a modification may significantly increase the time of the data analysis process, so it should be implemented only if necessary for the data set to analyze.

The tests with simple iron lines demonstrate the accuracy of the calculation of the transfer function and the line broadening kernel. We have further tested the accuracy of the full reflection model, which includes more processes such as calling the local reflection table and convolution. Due to light bending effect, the emission angle ($\theta_{\rm e}$) of the reflection component from the disk does not equal to the inclination angle ($i$) of the observer, and it differs from site to site on the accretion disk. With simulations assuming $\theta_{\rm e}=i$, we neglect this effect and test only the accuracy of the convolution process. We find that the parameter N\_ENER\_CONV, which controls the number of energy bins for the convolution, should be increased to 524288 for data quality of X-IFU and LAD. With simulations assuming the correct $\theta_{\rm e}$, neither \texttt{relxill} or \texttt{relxill\_nk} could provide acceptable fittings. For the data quality of current instruments, this discrepancy is not significant, as the impact of angle-averaged reflection models on parameter estimation is minor \citep{2020MNRAS.498.3565T}. Although it requires more intense computational power, full reflection models should be improved to implement the correct emission angle to prepare for data quality promised by X-IFU and LAD.

Last, we want to point out that in this work we have only tested the accuracy of the relativistic calculations of \texttt{relxill} and \texttt{relxill\_nk}. We have not investigated the accuracy of the atomic calculations of the {\tt xillver} model. The public {\tt xillver} tables are calculated assuming a number of simplifications; e.g., constant vertical density of the disk, angle of the incident radiation fixed to 45~deg, etc. The impact of these simplifications on the analysis of high-quality data should also be investigated. How spectra are tabulated is also important. For example, current {\tt xillver} spectra are tabulated with an energy resolution around 20~eV at the iron line, which has certainly to be improved for future analyses of X-IFU data (with an expected energy resolution around 3~eV).


\section*{Acknowledgements}

This work was supported by the National Natural Science Foundation of China (NSFC), Grant No.~11973019, 12250610185, and 12261131497, and the Natural Science Foundation of Shanghai, Grant No. 22ZR1403400.
T.M. also acknowledges the support from the China Scholarship Council (CSC), Grant No.~2022GXZ005433.


\section*{Data Availability}

The data underlying this article will be shared on reasonable request to the corresponding author.


\bibliographystyle{aasjournal}
\bibliography{Cite}

\begin{thebibliography}{}
\expandafter\ifx\csname natexlab\endcsname\relax\def\natexlab#1{#1}\fi
\providecommand{\url}[1]{\href{#1}{#1}}
\providecommand{\dodoi}[1]{doi:~\href{http://doi.org/#1}{\nolinkurl{#1}}}
\providecommand{\doeprint}[1]{\href{http://ascl.net/#1}{\nolinkurl{http://ascl.net/#1}}}
\providecommand{\doarXiv}[1]{\href{https://arxiv.org/abs/#1}{\nolinkurl{https://arxiv.org/abs/#1}}}

\bibitem[{{Abbott} {et~al.}(2016){Abbott}, {Abbott}, {Abbott}, {Abernathy},
  {Acernese}, {Ackley}, {Adams}, {Adams}, {Addesso}, {Adhikari}, {Adya},
  {Affeldt}, {Agathos}, {Agatsuma}, {Aggarwal}, {Aguiar}, {Aiello}, {Ain},
  {Ajith}, {Allen}, {Allocca}, {Altin}, {Anderson}, {Anderson}, {Arai},
  {Arain}, {Araya}, {Arceneaux}, {Areeda}, {Arnaud}, {Arun}, {Ascenzi},
  {Ashton}, {Ast}, {Aston}, {Astone}, {Aufmuth}, {Aulbert}, {Babak}, {Bacon},
  {Bader}, {Baker}, {Baldaccini}, {Ballardin}, {Ballmer}, {Barayoga},
  {Barclay}, {Barish}, {Barker}, {Barone}, {Barr}, {Barsotti}, {Barsuglia},
  {Barta}, {Bartlett}, {Barton}, {Bartos}, {Bassiri}, {Basti}, {Batch},
  {Baune}, {Bavigadda}, {Bazzan}, {Behnke}, {Bejger}, {Belczynski}, {Bell},
  {Bell}, {Berger}, {Bergman}, {Bergmann}, {Berry}, {Bersanetti}, {Bertolini},
  {Betzwieser}, {Bhagwat}, {Bhandare}, {Bilenko}, {Billingsley}, {Birch},
  {Birney}, {Birnholtz}, {Biscans}, {Bisht}, {Bitossi}, {Biwer}, {Bizouard},
  {Blackburn}, {Blair}, {Blair}, {Blair}, {Bloemen}, {Bock}, {Bodiya}, {Boer},
  {Bogaert}, {Bogan}, {Bohe}, {Bojtos}, {Bond}, {Bondu}, {Bonnand}, {Boom},
  {Bork}, {Boschi}, {Bose}, {Bouffanais}, {Bozzi}, {Bradaschia}, {Brady},
  {Braginsky}, {Branchesi}, {Brau}, {Briant}, {Brillet}, {Brinkmann},
  {Brisson}, {Brockill}, {Brooks}, {Brown}, {Brown}, {Brown}, {Buchanan},
  {Buikema}, {Bulik}, {Bulten}, {Buonanno}, {Buskulic}, {Buy}, {Byer},
  {Cabero}, {Cadonati}, {Cagnoli}, {Cahillane}, {Bustillo}, {Callister},
  {Calloni}, {Camp}, {Cannon}, {Cao}, {Capano}, {Capocasa}, {Carbognani},
  {Caride}, {Casanueva Diaz}, {Casentini}, {Caudill}, {Cavagli{\`a}},
  {Cavalier}, {Cavalieri}, {Cella}, {Cepeda}, {Baiardi}, {Cerretani},
  {Cesarini}, {Chakraborty}, {Chalermsongsak}, {Chamberlin}, {Chan}, {Chao},
  {Charlton}, {Chassande-Mottin}, {Chen}, {Chen}, {Cheng}, {Chincarini},
  {Chiummo}, {Cho}, {Cho}, {Chow}, {Christensen}, {Chu}, {Chua}, {Chung},
  {Ciani}, {Clara}, {Clark}, {Cleva}, {Coccia}, {Cohadon}, {Colla}, {Collette},
  {Cominsky}, {Constancio}, {Conte}, {Conti}, {Cook}, {Corbitt}, {Cornish},
  {Corsi}, {Cortese}, {Costa}, {Coughlin}, {Coughlin}, {Coulon}, {Countryman},
  {Couvares}, {Cowan}, {Coward}, {Cowart}, {Coyne}, {Coyne}, {Craig},
  {Creighton}, {Creighton}, {Cripe}, {Crowder}, {Cruise}, {Cumming},
  {Cunningham}, {Cuoco}, {Dal Canton}, {Danilishin}, {D'Antonio}, {Danzmann},
  {Darman}, {Da Silva Costa}, {Dattilo}, {Dave}, {Daveloza}, {Davier},
  {Davies}, {Daw}, {Day}, {De}, {DeBra}, {Debreczeni}, {Degallaix}, {De
  Laurentis}, {Del{\'e}glise}, {Del Pozzo}, {Denker}, {Dent}, {Dereli},
  {Dergachev}, {DeRosa}, {De Rosa}, {DeSalvo}, {Dhurandhar}, {D{\'\i}az}, {Di
  Fiore}, {Di Giovanni}, {Di Lieto}, {Di Pace}, {Di Palma}, {Di Virgilio},
  {Dojcinoski}, {Dolique}, {Donovan}, {Dooley}, {Doravari}, {Douglas},
  {Downes}, {Drago}, {Drever}, {Driggers}, {Du}, {Ducrot}, {Dwyer}, {Edo},
  {Edwards}, {Effler}, {Eggenstein}, {Ehrens}, {Eichholz}, {Eikenberry},
  {Engels}, {Essick}, {Etzel}, {Evans}, {Evans}, {Everett}, {Factourovich},
  {Fafone}, {Fair}, {Fairhurst}, {Fan}, {Fang}, {Farinon}, {Farr}, {Farr},
  {Favata}, {Fays}, {Fehrmann}, {Fejer}, {Feldbaum}, {Ferrante}, {Ferreira},
  {Ferrini}, {Fidecaro}, {Finn}, {Fiori}, {Fiorucci}, {Fisher}, {Flaminio},
  {Fletcher}, {Fong}, {Fournier}, {Franco}, {Frasca}, {Frasconi}, {Frede},
  {Frei}, {Freise}, {Frey}, {Frey}, {Fricke}, {Fritschel}, {Frolov}, {Fulda},
  {Fyffe}, {Gabbard}, {Gair}, {Gammaitoni}, {Gaonkar}, {Garufi}, {Gatto},
  {Gaur}, {Gehrels}, {Gemme}, {Gendre}, {Genin}, {Gennai}, {George}, {Gergely},
  {Germain}, {Ghosh}, {Ghosh}, {Ghosh}, {Giaime}, {Giardina}, {Giazotto},
  {Gill}, {Glaefke}, {Gleason}, {Goetz}, {Goetz}, {Gondan}, {Gonz{\'a}lez},
  {Castro}, {Gopakumar}, {Gordon}, {Gorodetsky}, {Gossan}, {Gosselin},
  {Gouaty}, {Graef}, {Graff}, {Granata}, {Grant}, {Gras}, {Gray}, {Greco},
  {Green}, {Greenhalgh}, {Groot}, {Grote}, {Grunewald}, {Guidi}, {Guo},
  {Gupta}, {Gupta}, {Gushwa}, {Gustafson}, {Gustafson}, {Hacker}, {Hall},
  {Hall}, {Hammond}, {Haney}, {Hanke}, {Hanks}, {Hanna}, {Hannam}, {Hanson},
  {Hardwick}, {Harms}, {Harry}, {Harry}, {Hart}, {Hartman}, {Haster},
  {Haughian}, {Healy}, {Heefner}, {Heidmann}, {Heintze}, {Heinzel}, {Heitmann},
  {Hello}, {Hemming}, {Hendry}, {Heng}, {Hennig}, {Heptonstall}, {Heurs},
  {Hild}, {Hoak}, {Hodge}, {Hofman}, {Hollitt}, {Holt}, {Holz}, {Hopkins},
  {Hosken}, {Hough}, {Houston}, {Howell}, {Hu}, {Huang}, {Huerta}, {Huet},
  {Hughey}, {Husa}, {Huttner}, {Huynh-Dinh}, {Idrisy}, {Indik}, {Ingram},
  {Inta}, {Isa}, {Isac}, {Isi}, {Islas}, {Isogai}, {Iyer}, {Izumi}, {Jacobson},
  {Jacqmin}, {Jang}, {Jani}, {Jaranowski}, {Jawahar}, {Jim{\'e}nez-Forteza},
  {Johnson}, {Johnson-McDaniel}, {Jones}, {Jones}, {Jonker}, {Ju}, {Haris},
  {Kalaghatgi}, {Kalogera}, {Kandhasamy}, {Kang}, {Kanner}, {Karki},
  {Kasprzack}, {Katsavounidis}, {Katzman}, {Kaufer}, {Kaur}, {Kawabe},
  {Kawazoe}, {K{\'e}f{\'e}lian}, {Kehl}, {Keitel}, {Kelley}, {Kells},
  {Kennedy}, {Keppel}, {Key}, {Khalaidovski}, {Khalili}, {Khan}, {Khan},
  {Khan}, {Khazanov}, {Kijbunchoo}, {Kim}, {Kim}, {Kim}, {Kim}, {Kim}, {Kim},
  {King}, {King}, {Kinzel}, {Kissel}, {Kleybolte}, {Klimenko}, {Koehlenbeck},
  {Kokeyama}, {Koley}, {Kondrashov}, {Kontos}, {Koranda}, {Korobko}, {Korth},
  {Kowalska}, {Kozak}, {Kringel}, {Krishnan}, {Kr{\'o}lak}, {Krueger}, {Kuehn},
  {Kumar}, {Kumar}, {Kuo}, {Kutynia}, {Kwee}, {Lackey}, {Landry}, {Lange},
  {Lantz}, {Lasky}, {Lazzarini}, {Lazzaro}, {Leaci}, {Leavey}, {Lebigot},
  {Lee}, {Lee}, {Lee}, {Lee}, {Lenon}, {Leonardi}, {Leong}, {Leroy},
  {Letendre}, {Levin}, {Levine}, {Li}, {Libson}, {Littenberg}, {Lockerbie},
  {Logue}, {Lombardi}, {London}, {Lord}, {Lorenzini}, {Loriette}, {Lormand},
  {Losurdo}, {Lough}, {Lousto}, {Lovelace}, {L{\"u}ck}, {Lundgren}, {Luo},
  {Lynch}, {Ma}, {MacDonald}, {Machenschalk}, {MacInnis}, {Macleod},
  {Maga{\~n}a-Sandoval}, {Magee}, {Mageswaran}, {Majorana}, {Maksimovic},
  {Malvezzi}, {Man}, {Mandel}, {Mandic}, {Mangano}, {Mansell}, {Manske},
  {Mantovani}, {Marchesoni}, {Marion}, {M{\'a}rka}, {M{\'a}rka}, {Markosyan},
  {Maros}, {Martelli}, {Martellini}, {Martin}, {Martin}, {Martynov}, {Marx},
  {Mason}, {Masserot}, {Massinger}, {Masso-Reid}, {Matichard}, {Matone},
  {Mavalvala}, {Mazumder}, {Mazzolo}, {McCarthy}, {McClelland}, {McCormick},
  {McGuire}, {McIntyre}, {McIver}, {McManus}, {McWilliams}, {Meacher},
  {Meadors}, {Meidam}, {Melatos}, {Mendell}, {Mendoza-Gandara}, {Mercer},
  {Merilh}, {Merzougui}, {Meshkov}, {Messenger}, {Messick}, {Meyers},
  {Mezzani}, {Miao}, {Michel}, {Middleton}, {Mikhailov}, {Milano}, {Miller},
  {Millhouse}, {Minenkov}, {Ming}, {Mirshekari}, {Mishra}, {Mitra},
  {Mitrofanov}, {Mitselmakher}, {Mittleman}, {Moggi}, {Mohan}, {Mohapatra},
  {Montani}, {Moore}, {Moore}, {Moraru}, {Moreno}, {Morriss}, {Mossavi},
  {Mours}, {Mow-Lowry}, {Mueller}, {Mueller}, {Muir}, {Mukherjee}, {Mukherjee},
  {Mukherjee}, {Mukund}, {Mullavey}, {Munch}, {Murphy}, {Murray}, {Mytidis},
  {Nardecchia}, {Naticchioni}, {Nayak}, {Necula}, {Nedkova}, {Nelemans},
  {Neri}, {Neunzert}, {Newton}, {Nguyen}, {Nielsen}, {Nissanke}, {Nitz},
  {Nocera}, {Nolting}, {Normandin}, {Nuttall}, {Oberling}, {Ochsner}, {O'Dell},
  {Oelker}, {Ogin}, {Oh}, {Oh}, {Ohme}, {Oliver}, {Oppermann}, {Oram},
  {O'Reilly}, {O'Shaughnessy}, {Ott}, {Ottaway}, {Ottens}, {Overmier}, {Owen},
  {Pai}, {Pai}, {Palamos}, {Palashov}, {Palomba}, {Pal-Singh}, {Pan}, {Pan},
  {Pankow}, {Pannarale}, {Pant}, {Paoletti}, {Paoli}, {Papa}, {Paris},
  {Parker}, {Pascucci}, {Pasqualetti}, {Passaquieti}, {Passuello},
  {Patricelli}, {Patrick}, {Pearlstone}, {Pedraza}, {Pedurand}, {Pekowsky},
  {Pele}, {Penn}, {Perreca}, {Pfeiffer}, {Phelps}, {Piccinni}, {Pichot},
  {Pickenpack}, {Piergiovanni}, {Pierro}, {Pillant}, {Pinard}, {Pinto},
  {Pitkin}, {Poeld}, {Poggiani}, {Popolizio}, {Post}, {Powell}, {Prasad},
  {Predoi}, {Premachandra}, {Prestegard}, {Price}, {Prijatelj}, {Principe},
  {Privitera}, {Prix}, {Prodi}, {Prokhorov}, {Puncken}, {Punturo}, {Puppo},
  {P{\"u}rrer}, {Qi}, {Qin}, {Quetschke}, {Quintero}, {Quitzow-James}, {Raab},
  {Rabeling}, {Radkins}, {Raffai}, {Raja}, {Rakhmanov}, {Ramet}, {Rapagnani},
  {Raymond}, {Razzano}, {Re}, {Read}, {Reed}, {Regimbau}, {Rei}, {Reid},
  {Reitze}, {Rew}, {Reyes}, {Ricci}, {Riles}, {Robertson}, {Robie}, {Robinet},
  {Rocchi}, {Rolland}, {Rollins}, {Roma}, {Romano}, {Romano}, {Romanov},
  {Romie}, {Rosi{\'n}ska}, {Rowan}, {R{\"u}diger}, {Ruggi}, {Ryan}, {Sachdev},
  {Sadecki}, {Sadeghian}, {Salconi}, {Saleem}, {Salemi}, {Samajdar}, {Sammut},
  {Sampson}, {Sanchez}, {Sandberg}, {Sandeen}, {Sanders}, {Sanders},
  {Sassolas}, {Sathyaprakash}, {Saulson}, {Sauter}, {Savage}, {Sawadsky},
  {Schale}, {Schilling}, {Schmidt}, {Schmidt}, {Schnabel}, {Schofield},
  {Sch{\"o}nbeck}, {Schreiber}, {Schuette}, {Schutz}, {Scott}, {Scott},
  {Sellers}, {Sengupta}, {Sentenac}, {Sequino}, {Sergeev}, {Serna},
  {Setyawati}, {Sevigny}, {Shaddock}, {Shaffer}, {Shah}, {Shahriar}, {Shaltev},
  {Shao}, {Shapiro}, {Shawhan}, {Sheperd}, {Shoemaker}, {Shoemaker}, {Siellez},
  {Siemens}, {Sigg}, {Silva}, {Simakov}, {Singer}, {Singer}, {Singh}, {Singh},
  {Singhal}, {Sintes}, {Slagmolen}, {Smith}, {Smith}, {Smith}, {Smith}, {Son},
  {Sorazu}, {Sorrentino}, {Souradeep}, {Srivastava}, {Staley}, {Steinke},
  {Steinlechner}, {Steinlechner}, {Steinmeyer}, {Stephens}, {Stevenson},
  {Stone}, {Strain}, {Straniero}, {Stratta}, {Strauss}, {Strigin}, {Sturani},
  {Stuver}, {Summerscales}, {Sun}, {Sutton}, {Swinkels}, {Szczepa{\'n}czyk},
  {Tacca}, {Talukder}, {Tanner}, {T{\'a}pai}, {Tarabrin}, {Taracchini},
  {Taylor}, {Theeg}, {Thirugnanasambandam}, {Thomas}, {Thomas}, {Thomas},
  {Thorne}, {Thorne}, {Thrane}, {Tiwari}, {Tiwari}, {Tokmakov}, {Tomlinson},
  {Tonelli}, {Torres}, {Torrie}, {T{\"o}yr{\"a}}, {Travasso}, {Traylor},
  {Trifir{\`o}}, {Tringali}, {Trozzo}, {Tse}, {Turconi}, {Tuyenbayev},
  {Ugolini}, {Unnikrishnan}, {Urban}, {Usman}, {Vahlbruch}, {Vajente},
  {Valdes}, {Vallisneri}, {van Bakel}, {van Beuzekom}, {van den Brand}, {Van
  Den Broeck}, {Vander-Hyde}, {van der Schaaf}, {van Heijningen}, {van Veggel},
  {Vardaro}, {Vass}, {Vas{\'u}th}, {Vaulin}, {Vecchio}, {Vedovato}, {Veitch},
  {Veitch}, {Venkateswara}, {Verkindt}, {Vetrano}, {Vicer{\'e}}, {Vinciguerra},
  {Vine}, {Vinet}, {Vitale}, {Vo}, {Vocca}, {Vorvick}, {Voss}, {Vousden},
  {Vyatchanin}, {Wade}, {Wade}, {Wade}, {Waldman}, {Walker}, {Wallace},
  {Walsh}, {Wang}, {Wang}, {Wang}, {Wang}, {Wang}, {Ward}, {Ward}, {Warner},
  {Was}, {Weaver}, {Wei}, {Weinert}, {Weinstein}, {Weiss}, {Welborn}, {Wen},
  {We{\ss}els}, {Westphal}, {Wette}, {Whelan}, {Whitcomb}, {White}, {Whiting},
  {Wiesner}, {Wilkinson}, {Willems}, {Williams}, {Williams}, {Williamson},
  {Willis}, {Willke}, {Wimmer}, {Winkelmann}, {Winkler}, {Wipf}, {Wiseman},
  {Wittel}, {Woan}, {Worden}, {Wright}, {Wu}, {Yablon}, {Yakushin}, {Yam},
  {Yamamoto}, {Yancey}, {Yap}, {Yu}, {Yvert}, {Zadro{\.Z}ny}, {Zangrando},
  {Zanolin}, {Zendri}, {Zevin}, {Zhang}, {Zhang}, {Zhang}, {Zhang}, {Zhao},
  {Zhou}, {Zhou}, {Zhu}, {Zucker}, {Zuraw}, {Zweizig}, {LIGO Scientific
  Collaboration}, \& {Virgo Collaboration}}]{2016PhRvL.116f1102A}
{Abbott}, B.~P., {Abbott}, R., {Abbott}, T.~D., {et~al.} 2016, \prl, 116,
  061102, \dodoi{10.1103/PhysRevLett.116.061102}

\bibitem[{{Abdikamalov} {et~al.}(2019){Abdikamalov}, {Ayzenberg}, {Bambi},
  {Dauser}, {Garc{\'\i}a}, \& {Nampalliwar}}]{2019ApJ...878...91A}
{Abdikamalov}, A.~B., {Ayzenberg}, D., {Bambi}, C., {et~al.} 2019, \apj, 878,
  91, \dodoi{10.3847/1538-4357/ab1f89}

\bibitem[{{Abdikamalov} {et~al.}(2024){Abdikamalov}, {Ayzenberg}, {Bambi},
  {Mirzaev}, {Riaz}, \& {Shashank}}]{zenodo24}
---. 2024, Zenodo, \dodoi{10.5281/zenodo.10673858}

\bibitem[{{Arnaud}(1996)}]{1996ASPC..101...17A}
{Arnaud}, K.~A. 1996, in Astronomical Society of the Pacific Conference Series,
  Vol. 101, Astronomical Data Analysis Software and Systems V, ed. G.~H.
  {Jacoby} \& J.~{Barnes}, 17

\bibitem[{{Bambi}(2013)}]{Bambi2013}
{Bambi}, C. 2013, \prd, 87, 023007, \dodoi{10.1103/PhysRevD.87.023007}

\bibitem[{{Bambi}(2017)}]{2017bhlt.book.....B}
---. 2017, {Black Holes: A Laboratory for Testing Strong Gravity} (Springer
  Singapore), \dodoi{10.1007/978-981-10-4524-0}

\bibitem[{{Bambi}(2022)}]{2022arXiv221005322B}
---. 2022, arXiv e-prints, arXiv:2210.05322, \dodoi{10.48550/arXiv.2210.05322}

\bibitem[{{Bambi} {et~al.}(2017){Bambi}, {C{\'a}rdenas-Avenda{\~n}o}, {Dauser},
  {Garc{\'\i}a}, \& {Nampalliwar}}]{Bambi_2017}
{Bambi}, C., {C{\'a}rdenas-Avenda{\~n}o}, A., {Dauser}, T., {Garc{\'\i}a},
  J.~A., \& {Nampalliwar}, S. 2017, \apj, 842, 76,
  \dodoi{10.3847/1538-4357/aa74c0}

\bibitem[{{Bambi} {et~al.}(2021){Bambi}, {Brenneman}, {Dauser}, {Garc{\'\i}a},
  {Grinberg}, {Ingram}, {Jiang}, {Liu}, {Lohfink}, {Marinucci}, {Mastroserio},
  {Middei}, {Nampalliwar}, {Nied{\'z}wiecki}, {Steiner}, {Tripathi}, \&
  {Zdziarski}}]{2021SSRv..217...65B}
{Bambi}, C., {Brenneman}, L.~W., {Dauser}, T., {et~al.} 2021, \ssr, 217, 65,
  \dodoi{10.1007/s11214-021-00841-8}

\bibitem[{{Cao} {et~al.}(2018){Cao}, {Nampalliwar}, {Bambi}, {Dauser}, \&
  {Garc{\'\i}a}}]{2018PhRvL.120e1101C}
{Cao}, Z., {Nampalliwar}, S., {Bambi}, C., {Dauser}, T., \& {Garc{\'\i}a},
  J.~A. 2018, \prl, 120, 051101, \dodoi{10.1103/PhysRevLett.120.051101}

\bibitem[{{Connors} {et~al.}(2021){Connors}, {Steiner}, {Homan}, {Garcia},
  {Mastroserio}, {Gendreau}, \& {Arzoumanian}}]{Connors2021ATel14725....1C}
{Connors}, R., {Steiner}, J., {Homan}, J., {et~al.} 2021, The Astronomer's
  Telegram, 14725, 1

\bibitem[{{Cunningham}(1975)}]{1975ApJ...202..788C}
{Cunningham}, C.~T. 1975, \apj, 202, 788, \dodoi{10.1086/154033}

\bibitem[{{Dauser} {et~al.}(2013){Dauser}, {Garcia}, {Wilms}, {B{\"o}ck},
  {Brenneman}, {Falanga}, {Fukumura}, \& {Reynolds}}]{2013MNRAS.430.1694D}
{Dauser}, T., {Garcia}, J., {Wilms}, J., {et~al.} 2013, \mnras, 430, 1694,
  \dodoi{10.1093/mnras/sts710}

\bibitem[{{Dauser} {et~al.}(2010){Dauser}, {Wilms}, {Reynolds}, \&
  {Brenneman}}]{Dauser2010MNRAS.409.1534D}
{Dauser}, T., {Wilms}, J., {Reynolds}, C.~S., \& {Brenneman}, L.~W. 2010,
  \mnras, 409, 1534, \dodoi{10.1111/j.1365-2966.2010.17393.x}

\bibitem[{{Draghis} {et~al.}(2023){Draghis}, {Miller}, {Zoghbi}, {Reynolds},
  {Costantini}, {Gallo}, \& {Tomsick}}]{2023ApJ...946...19D}
{Draghis}, P.~A., {Miller}, J.~M., {Zoghbi}, A., {et~al.} 2023, \apj, 946, 19,
  \dodoi{10.3847/1538-4357/acafe7}

\bibitem[{{Fabian} {et~al.}(1995){Fabian}, {Nandra}, {Reynolds}, {Brandt},
  {Otani}, {Tanaka}, {Inoue}, \& {Iwasawa}}]{1995MNRAS.277L..11F}
{Fabian}, A.~C., {Nandra}, K., {Reynolds}, C.~S., {et~al.} 1995, \mnras, 277,
  L11, \dodoi{10.1093/mnras/277.1.L11}

\bibitem[{{Fabian} {et~al.}(1989){Fabian}, {Rees}, {Stella}, \&
  {White}}]{1989MNRAS.238..729F}
{Fabian}, A.~C., {Rees}, M.~J., {Stella}, L., \& {White}, N.~E. 1989, \mnras,
  238, 729, \dodoi{10.1093/mnras/238.3.729}

\bibitem[{{Garc{\'\i}a} {et~al.}(2013){Garc{\'\i}a}, {Dauser}, {Reynolds},
  {Kallman}, {McClintock}, {Wilms}, \& {Eikmann}}]{2013ApJ...768..146G}
{Garc{\'\i}a}, J., {Dauser}, T., {Reynolds}, C.~S., {et~al.} 2013, \apj, 768,
  146, \dodoi{10.1088/0004-637X/768/2/146}

\bibitem[{{Garc{\'\i}a} \& {Kallman}(2010)}]{2010ApJ...718..695G}
{Garc{\'\i}a}, J., \& {Kallman}, T.~R. 2010, \apj, 718, 695,
  \dodoi{10.1088/0004-637X/718/2/695}

\bibitem[{{Garc{\'\i}a} {et~al.}(2014){Garc{\'\i}a}, {Dauser}, {Lohfink},
  {Kallman}, {Steiner}, {McClintock}, {Brenneman}, {Wilms}, {Eikmann},
  {Reynolds}, \& {Tombesi}}]{2014ApJ...782...76G}
{Garc{\'\i}a}, J., {Dauser}, T., {Lohfink}, A., {et~al.} 2014, \apj, 782, 76,
  \dodoi{10.1088/0004-637X/782/2/76}

\bibitem[{{Ingram} {et~al.}(2019){Ingram}, {Mastroserio}, {Dauser},
  {Hovenkamp}, {van der Klis}, \& {Garc{\'\i}a}}]{2019MNRAS.488..324I}
{Ingram}, A., {Mastroserio}, G., {Dauser}, T., {et~al.} 2019, \mnras, 488, 324,
  \dodoi{10.1093/mnras/stz1720}

\bibitem[{{Jiang} {et~al.}(2018){Jiang}, {Parker}, {Fabian}, {Alston},
  {Buisson}, {Cackett}, {Chiang}, {Dauser}, {Gallo}, {Garc{\'\i}a}, {Harrison},
  {Lohfink}, {De Marco}, {Kara}, {Miller}, {Miniutti}, {Pinto}, {Walton}, \&
  {Wilkins}}]{Jiang2018MNRAS.477.3711J}
{Jiang}, J., {Parker}, M.~L., {Fabian}, A.~C., {et~al.} 2018, \mnras, 477,
  3711, \dodoi{10.1093/mnras/sty836}

\bibitem[{{Johannsen}(2013)}]{2013PhRvD..88d4002J}
{Johannsen}, T. 2013, \prd, 88, 044002, \dodoi{10.1103/PhysRevD.88.044002}

\bibitem[{{Kaastra} \& {Bleeker}(2016)}]{Kaastra2016A&A...587A.151K}
{Kaastra}, J.~S., \& {Bleeker}, J.~A.~M. 2016, Astron. Astrophys., 587, A151,
  \dodoi{10.1051/0004-6361/201527395}

\bibitem[{{Laor}(1991)}]{1991ApJ...376...90L}
{Laor}, A. 1991, \apj, 376, 90, \dodoi{10.1086/170257}

\bibitem[{{Liu} {et~al.}(2022{\natexlab{a}}){Liu}, {Jiang}, {Zhang}, {Bambi},
  {Ji}, {Kong}, \& {Zhang}}]{Liu2022MNRAS.513.4308L}
{Liu}, H., {Jiang}, J., {Zhang}, Z., {et~al.} 2022{\natexlab{a}}, \mnras, 513,
  4308, \dodoi{10.1093/mnras/stac1178}

\bibitem[{{Liu} {et~al.}(2022{\natexlab{b}}){Liu}, {Bambi}, {Jiang}, {Garcia},
  {Ji}, {Kong}, {Ren}, {Zhang}, \& {Zhang}}]{Liu2022arXiv221109543L}
{Liu}, H., {Bambi}, C., {Jiang}, J., {et~al.} 2022{\natexlab{b}}, arXiv
  e-prints, arXiv:2211.09543, \dodoi{10.48550/arXiv.2211.09543}

\bibitem[{{McClintock} {et~al.}(2014){McClintock}, {Narayan}, \&
  {Steiner}}]{2014SSRv..183..295M}
{McClintock}, J.~E., {Narayan}, R., \& {Steiner}, J.~F. 2014, \ssr, 183, 295,
  \dodoi{10.1007/s11214-013-0003-9}

\bibitem[{{Miller} {et~al.}(2009){Miller}, {Reynolds}, {Fabian}, {Miniutti}, \&
  {Gallo}}]{2009ApJ...697..900M}
{Miller}, J.~M., {Reynolds}, C.~S., {Fabian}, A.~C., {Miniutti}, G., \&
  {Gallo}, L.~C. 2009, \apj, 697, 900, \dodoi{10.1088/0004-637X/697/1/900}

\bibitem[{{Nandra} {et~al.}(2007){Nandra}, {O'Neill}, {George}, \&
  {Reeves}}]{2007MNRAS.382..194N}
{Nandra}, K., {O'Neill}, P.~M., {George}, I.~M., \& {Reeves}, J.~N. 2007,
  \mnras, 382, 194, \dodoi{10.1111/j.1365-2966.2007.12331.x}

\bibitem[{{Nandra} {et~al.}(2013){Nandra}, {Barret}, {Barcons}, {Fabian}, {den
  Herder}, {Piro}, {Watson}, {Adami}, {Aird}, {Afonso}, {Alexander},
  {Argiroffi}, {Amati}, {Arnaud}, {Atteia}, {Audard}, {Badenes}, {Ballet},
  {Ballo}, {Bamba}, {Bhardwaj}, {Stefano Battistelli}, {Becker}, {De Becker},
  {Behar}, {Bianchi}, {Biffi}, {B{\^\i}rzan}, {Bocchino}, {Bogdanov}, {Boirin},
  {Boller}, {Borgani}, {Borm}, {Bouch{\'e}}, {Bourdin}, {Bower}, {Braito},
  {Branchini}, {Branduardi-Raymont}, {Bregman}, {Brenneman}, {Brightman},
  {Br{\"u}ggen}, {Buchner}, {Bulbul}, {Brusa}, {Bursa}, {Caccianiga},
  {Cackett}, {Campana}, {Cappelluti}, {Cappi}, {Carrera}, {Ceballos},
  {Christensen}, {Chu}, {Churazov}, {Clerc}, {Corbel}, {Corral}, {Comastri},
  {Costantini}, {Croston}, {Dadina}, {D'Ai}, {Decourchelle}, {Della Ceca},
  {Dennerl}, {Dolag}, {Done}, {Dovciak}, {Drake}, {Eckert}, {Edge}, {Ettori},
  {Ezoe}, {Feigelson}, {Fender}, {Feruglio}, {Finoguenov}, {Fiore}, {Galeazzi},
  {Gallagher}, {Gandhi}, {Gaspari}, {Gastaldello}, {Georgakakis},
  {Georgantopoulos}, {Gilfanov}, {Gitti}, {Gladstone}, {Goosmann}, {Gosset},
  {Grosso}, {Guedel}, {Guerrero}, {Haberl}, {Hardcastle}, {Heinz}, {Alonso
  Herrero}, {Herv{\'e}}, {Holmstrom}, {Iwasawa}, {Jonker}, {Kaastra}, {Kara},
  {Karas}, {Kastner}, {King}, {Kosenko}, {Koutroumpa}, {Kraft}, {Kreykenbohm},
  {Lallement}, {Lanzuisi}, {Lee}, {Lemoine-Goumard}, {Lobban}, {Lodato},
  {Lovisari}, {Lotti}, {McCharthy}, {McNamara}, {Maggio}, {Maiolino}, {De
  Marco}, {de Martino}, {Mateos}, {Matt}, {Maughan}, {Mazzotta}, {Mendez},
  {Merloni}, {Micela}, {Miceli}, {Mignani}, {Miller}, {Miniutti}, {Molendi},
  {Montez}, {Moretti}, {Motch}, {Naz{\'e}}, {Nevalainen}, {Nicastro}, {Nulsen},
  {Ohashi}, {O'Brien}, {Osborne}, {Oskinova}, {Pacaud}, {Paerels}, {Page},
  {Papadakis}, {Pareschi}, {Petre}, {Petrucci}, {Piconcelli}, {Pillitteri},
  {Pinto}, {de Plaa}, {Pointecouteau}, {Ponman}, {Ponti}, {Porquet}, {Pounds},
  {Pratt}, {Predehl}, {Proga}, {Psaltis}, {Rafferty}, {Ramos-Ceja}, {Ranalli},
  {Rasia}, {Rau}, {Rauw}, {Rea}, {Read}, {Reeves}, {Reiprich}, {Renaud},
  {Reynolds}, {Risaliti}, {Rodriguez}, {Rodriguez Hidalgo}, {Roncarelli},
  {Rosario}, {Rossetti}, {Rozanska}, {Rovilos}, {Salvaterra}, {Salvato}, {Di
  Salvo}, {Sanders}, {Sanz-Forcada}, {Schawinski}, {Schaye}, {Schwope},
  {Sciortino}, {Severgnini}, {Shankar}, {Sijacki}, {Sim}, {Schmid}, {Smith},
  {Steiner}, {Stelzer}, {Stewart}, {Strohmayer}, {Str{\"u}der}, {Sun}, {Takei},
  {Tatischeff}, {Tiengo}, {Tombesi}, {Trinchieri}, {Tsuru}, {Ud-Doula},
  {Ursino}, {Valencic}, {Vanzella}, {Vaughan}, {Vignali}, {Vink}, {Vito},
  {Volonteri}, {Wang}, {Webb}, {Willingale}, {Wilms}, {Wise}, {Worrall},
  {Young}, {Zampieri}, {In't Zand}, {Zane}, {Zezas}, {Zhang}, \&
  {Zhuravleva}}]{2013arXiv1306.2307N}
{Nandra}, K., {Barret}, D., {Barcons}, X., {et~al.} 2013, arXiv e-prints,
  arXiv:1306.2307, \dodoi{10.48550/arXiv.1306.2307}

\bibitem[{{Riaz} {et~al.}(2022){Riaz}, {Abdikamalov}, {Ayzenberg}, {Bambi},
  {Wang}, \& {Yu}}]{2022ApJ...925...51R}
{Riaz}, S., {Abdikamalov}, A.~B., {Ayzenberg}, D., {et~al.} 2022, \apj, 925,
  51, \dodoi{10.3847/1538-4357/ac3827}

\bibitem[{{Ross} \& {Fabian}(2005)}]{2005MNRAS.358..211R}
{Ross}, R.~R., \& {Fabian}, A.~C. 2005, \mnras, 358, 211,
  \dodoi{10.1111/j.1365-2966.2005.08797.x}

\bibitem[{{Shidatsu} {et~al.}(2019){Shidatsu}, {Nakahira}, {Murata}, {Adachi},
  {Kawai}, {Ueda}, \& {Negoro}}]{Shidatsu2019ApJ...874..183S}
{Shidatsu}, M., {Nakahira}, S., {Murata}, K.~L., {et~al.} 2019, \apj, 874, 183,
  \dodoi{10.3847/1538-4357/ab09ff}

\bibitem[{{Tanaka} {et~al.}(1995){Tanaka}, {Nandra}, {Fabian}, {Inoue},
  {Otani}, {Dotani}, {Hayashida}, {Iwasawa}, {Kii}, {Kunieda}, {Makino}, \&
  {Matsuoka}}]{1995Natur.375..659T}
{Tanaka}, Y., {Nandra}, K., {Fabian}, A.~C., {et~al.} 1995, \nat, 375, 659,
  \dodoi{10.1038/375659a0}

\bibitem[{{Tripathi} {et~al.}(2020){Tripathi}, {Liu}, \&
  {Bambi}}]{2020MNRAS.498.3565T}
{Tripathi}, A., {Liu}, H., \& {Bambi}, C. 2020, \mnras, 498, 3565,
  \dodoi{10.1093/mnras/staa2618}

\bibitem[{{Tripathi} {et~al.}(2019){Tripathi}, {Nampalliwar}, {Abdikamalov},
  {Ayzenberg}, {Bambi}, {Dauser}, {Garc{\'\i}a}, \&
  {Marinucci}}]{2019ApJ...875...56T}
{Tripathi}, A., {Nampalliwar}, S., {Abdikamalov}, A.~B., {et~al.} 2019, \apj,
  875, 56, \dodoi{10.3847/1538-4357/ab0e7e}

\bibitem[{{Tripathi} {et~al.}(2021){Tripathi}, {Zhang}, {Abdikamalov},
  {Ayzenberg}, {Bambi}, {Jiang}, {Liu}, \& {Zhou}}]{2021ApJ...913...79T}
{Tripathi}, A., {Zhang}, Y., {Abdikamalov}, A.~B., {et~al.} 2021, \apj, 913,
  79, \dodoi{10.3847/1538-4357/abf6cd}

\bibitem[{{Vitale} {et~al.}(2014){Vitale}, {Lynch}, {Veitch}, {Raymond}, \&
  {Sturani}}]{2014PhRvL.112y1101V}
{Vitale}, S., {Lynch}, R., {Veitch}, J., {Raymond}, V., \& {Sturani}, R. 2014,
  \prl, 112, 251101, \dodoi{10.1103/PhysRevLett.112.251101}

\bibitem[{{Zhang} {et~al.}(1997){Zhang}, {Cui}, \&
  {Chen}}]{1997ApJ...482L.155Z}
{Zhang}, S.~N., {Cui}, W., \& {Chen}, W. 1997, Astrophys. J. Lett., 482, L155,
  \dodoi{10.1086/310705}

\bibitem[{{Zhang} {et~al.}(2016){Zhang}, {Feroci}, {Santangelo}, {Dong},
  {Feng}, {Lu}, {Nandra}, {Wang}, {Zhang}, {Bozzo}, {Brandt}, {De Rosa}, {Gou},
  {Hernanz}, {van der Klis}, {Li}, {Liu}, {Orleanski}, {Pareschi}, {Pohl},
  {Poutanen}, {Qu}, {Schanne}, {Stella}, {Uttley}, {Watts}, {Xu}, {Yu}, {in 't
  Zand}, {Zane}, {Alvarez}, {Amati}, {Baldini}, {Bambi}, {Basso},
  {Bhattacharyya S.}, {}, {Belloni}, {Bellutti}, {Bianchi}, {Brez}, {Bursa},
  {Burwitz}, {Budtz-J{\o}rgensen}, {Caiazzo}, {Campana}, {Cao}, {Casella},
  {Chen}, {Chen}, {Chen}, {Chen}, {Chen}, {Chen}, {Civitani}, {Coti Zelati},
  {Cui}, {Cui}, {Dai}, {Del Monte}, {de Martino}, {Di Cosimo}, {Diebold},
  {Dovciak}, {Donnarumma}, {Doroshenko}, {Esposito}, {Evangelista}, {Favre},
  {Friedrich}, {Fuschino}, {Galvez}, {Gao}, {Ge}, {Gevin}, {Goetz}, {Han},
  {Heyl}, {Horak}, {Hu}, {Huang}, {Huang}, {Hudec}, {Huppenkothen}, {Israel},
  {Ingram}, {Karas}, {Karelin}, {Jenke}, {Ji}, {Korpela}, {Kunneriath},
  {Labanti}, {Li}, {Li}, {Li}, {Liang}, {Limousin}, {Lin}, {Ling}, {Liu},
  {Liu}, {Liu}, {Lu}, {Lund}, {Lai}, {Luo}, {Luo}, {Ma}, {Mahmoodifar},
  {Marisaldi}, {Martindale}, {Meidinger}, {Men}, {Michalska}, {Mignani},
  {Minuti}, {Motta}, {Muleri}, {Neilsen}, {Orlandini}, {Pan}, {Patruno},
  {Perinati}, {Picciotto}, {Piemonte}, {Pinchera}, {Rachevski A.}, {Rapisarda},
  {Rea}, {Rossi}, {Rubini}, {Sala}, {Shu}, {Sgro}, {Shen}, {Soffitta}, {Song},
  {Spandre}, {Stratta}, {Strohmayer}, {Sun}, {Svoboda}, {Tagliaferri},
  {Tenzer}, {Hong}, {Taverna}, {Torok}, {Turolla}, {Vacchi}, {Wang}, {Walton},
  {Wang}, {Wang}, {Wang}, {Wang}, {Weng}, {Wilms}, {Winter}, {Wu}, {Wu},
  {Xiong}, {Xu}, {Xue}, {Yan}, {Yang}, {Yang}, {Yang}, {Yuan}, {Yuan}, {Yuan},
  {Zampa}, {Zampa}, {Zdziarski}, {Zhang}, {Zhang}, {Zhang}, {Zhang}, {Zhang},
  {Zhang}, {Zheng}, {Zhou}, \& {Zhou X.~L.}}]{2016SPIE.9905E..1QZ}
{Zhang}, S.~N., {Feroci}, M., {Santangelo}, A., {et~al.} 2016, in Society of
  Photo-Optical Instrumentation Engineers (SPIE) Conference Series, Vol. 9905,
  Space Telescopes and Instrumentation 2016: Ultraviolet to Gamma Ray, ed.
  J.-W.~A. {den Herder}, T.~{Takahashi}, \& M.~{Bautz}, 99051Q,
  \dodoi{10.1117/12.2232034}

\bibitem[{{Zhang} {et~al.}(2022){Zhang}, {Liu}, {Abdikamalov}, {Ayzenberg},
  {Bambi}, \& {Zhou}}]{2022ApJ...924...72Z}
{Zhang}, Z., {Liu}, H., {Abdikamalov}, A.~B., {et~al.} 2022, \apj, 924, 72,
  \dodoi{10.3847/1538-4357/ac350e}

\end{thebibliography}

\end{document}